\documentclass[sigconf,screen]{acmart}

\usepackage[ruled]{algorithm2e}
\usepackage{color}
\usepackage[normalem]{ulem}
\usepackage{multirow}
\usepackage{subfigure}
\usepackage{sistyle}
\SIthousandsep{,}
\usepackage{makecell}
\usepackage{multirow}
\usepackage[inline]{enumitem}
\usepackage{filecontents}
\usepackage{bbm}
\usepackage{pgfplots}
\usetikzlibrary{patterns}
\usepackage[skip=0pt]{caption}

\usepackage{acronym}
\acrodef{IR}{information retrieval}
\acrodef{IL}{incremental learning}
\acrodef{CL}{continual learning}
\acrodef{GR}{generative retrieval}

\newcommand{\heading}[1]{\vspace*{1mm}\noindent\textbf{#1.}}

\AtBeginDocument{%
}

\makeatletter
\g@addto@macro\normalsize{%
  \abovedisplayskip 1pt plus1pt 
  \belowdisplayskip 1pt plus1pt
  \abovedisplayshortskip  1pt plus1pt%
  \belowdisplayshortskip  1pt plus1pt
}
\makeatother

\copyrightyear{2023}
\acmYear{2023}
\setcopyright{rightsretained}
\acmConference[CIKM '23]{Proceedings of the 32nd ACM International Conference on Information and Knowledge Management}{October 21--25, 2023}{Birmingham, United Kingdom}
\acmBooktitle{Proceedings of the 32nd ACM International Conference on Information and Knowledge Management (CIKM '23), October 21--25, 2023, Birmingham, United Kingdom}\acmDOI{10.1145/3583780.3614821}
\acmISBN{979-8-4007-0124-5/23/10}

\makeatletter
\gdef\@copyrightpermission{
 \begin{minipage}{0.3\columnwidth}
  \href{https://creativecommons.org/licenses/by/4.0/}{\includegraphics[width=0.90\textwidth]{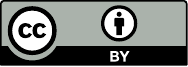}}
 \end{minipage}\hfill
 \begin{minipage}{0.7\columnwidth}
  \href{https://creativecommons.org/licenses/by/4.0/}{This work is licensed under a Creative Commons Attribution International 4.0 License.}
 \end{minipage}
 \vspace{5pt}
}
\makeatother

\begin{document}

\title[Continual Learning for Generative Retrieval over Dynamic Corpora]{Continual   Learning for Generative Retrieval\\ over  Dynamic Corpora}

\author{Jiangui Chen}
\orcid{0000-0002-6235-6526}
\author{Ruqing Zhang}
\orcid{0000-0003-4294-2541}
\authornote{Research conducted when the author was at the University of Amsterdam.}
\affiliation{
	\institution{CAS Key Lab of Network Data Science and Technology, ICT, CAS}
	\institution{University of Chinese Academy of Sciences}
	\city{Beijing}
	\country{China}
}
\email{{chenjiangui18z, zhangruqing}@ict.ac.cn}
 
 
\author{Jiafeng Guo}
\orcid{0000-0002-9509-8674}
\authornote{Jiafeng Guo is the corresponding author.}
\affiliation{
	\institution{CAS Key Lab of Network Data Science and Technology, ICT, CAS}
	\institution{University of Chinese Academy of Sciences}
	\city{Beijing}
	\country{China}
}
\email{guojiafeng@ict.ac.cn}

\author{Maarten de Rijke}
\orcid{0000-0002-1086-0202}
\affiliation{
 \institution{University of Amsterdam}
 \city{Amsterdam}
 \country{The Netherlands}
}
\email{m.derijke@uva.nl}

\author{Wei Chen}
\orcid{0000-0002-7438-5180}
\affiliation{
	\institution{CAS Key Lab of Network Data Science and Technology, ICT, CAS}
	\institution{University of Chinese Academy of Sciences}
 \city{Beijing}
 \country{China}
}
\email{chenwei2022@ict.ac.cn}

\author{Yixing Fan}
\orcid{0000-0003-4317-2702}
\affiliation{
	\institution{CAS Key Lab of Network Data Science and Technology, ICT, CAS}
	\institution{University of Chinese Academy of Sciences}
	\city{Beijing}
	\country{China}
}
\email{fanyixing@ict.ac.cn}
 
\author{Xueqi Cheng}
\orcid{0000-0002-5201-8195}
\affiliation{
	\institution{CAS Key Lab of Network Data Science and Technology, ICT, CAS}
	\institution{University of Chinese Academy of Sciences}
	\city{Beijing}
	\country{China}
}
\email{cxq@ict.ac.cn}

\renewcommand{\shortauthors}{Jiangui Chen et al.}

\begin{abstract}

\Acf{GR} directly predicts the identifiers of relevant documents (i.e., docids) based on a parametric model.
It has achieved solid performance on many ad-hoc retrieval tasks.  
So far, these tasks have assumed a static document collection. 
In many practical scenarios, however, document collections are dynamic, where new documents are continuously added to the corpus. 
The ability to incrementally index new documents while preserving the ability to answer queries with both previously and newly indexed relevant documents is vital to applying \ac{GR} models. 
In this paper, we address this practical continual learning problem for \ac{GR}. We put forward a novel Continual-LEarner for generatiVE Retrieval (CLEVER) model and make two major contributions to continual learning for \ac{GR}:  
\begin{enumerate*}[label=(\roman*)]
    \item To encode new documents into docids with low computational cost, we present Incremental Product Quantization, which updates a partial quantization codebook according to two adaptive thresholds; and 
    \item To memorize new documents for querying without forgetting previous knowledge, we propose a memory-augmented learning mechanism, to form meaningful connections between old and new documents. 
\end{enumerate*}
Empirical results demonstrate the effectiveness and efficiency of the proposed model.  
\end{abstract}



\maketitle

\acresetall

\vspace{-2mm}
\section{Introduction}

\Acf{GR} has emerged as a new paradigm for \acf{IR}~\cite{metzler2021rethinking}. 
Without loss of generality, the \ac{GR} paradigm aims to integrate all necessary relevant information in the collection into a single,  consolidated model. 
With \ac{GR}, indexing is replaced by model training, while retrieval is replaced by model inference.
A sequence-to-sequence (seq2seq) model is jointly trained for both indexing and retrieval tasks: the indexing task aims to associate the document content with its identifiers (i.e., docids); the retrieval task requires that queries are mapped to relevant docids.

\heading{\ac{GR} and dynamic corpora}
Most existing work on GR assumes a stationary learning scenario, i.e., the document collection is fixed \cite{dsi,nci,genre,corpusbrain}.  
However, dynamic corpora are a common setting for \ac{IR}.
In most real-world scenarios, information changes and new documents emerge incrementally over time. 
For example, in digital libraries, new electronic collections are continuously added to the system~\citep{witten-2010-how}. 
And a medical search engine may continuously expand its coverage to provide information about emerging diseases, as we have seen with COVID-19~\citep{kousha-2020-covid-19}. 
An important difference between static and dynamic scenarios is that in the former scenario a \ac{GR} system may be provided with abundant labels for training, but in the latter scenario very few labeled query-document pairs are typically available. 
Therefore, it is critical to study the continual learning ability of \ac{GR} models before their use in real-world environments. 

The \emph{continual document learning} task comes with interesting challenges. 
For traditional pipeline frameworks for IR \cite{bm25, guo2016deep, dpr}, indexing and retrieval are two separate modules. 
Therefore, when new documents arrive, their encoded representations can be directly included in an external index without updating the retrieval model due to the decoupled architecture. 
In \ac{GR}, all document information is encoded into the model parameters. 
To add new documents to the internal index (i.e., model parameters), the GR model must be re-trained from scratch every time the underlying corpus is updated. 
Clearly, due to the high computational costs, this is not a feasible way of handling a dynamically evolving document collection.

\heading{A document-incremental retriever}
Our aim is to develop an effective and efficient \emph{Continual-LEarner for generatiVE Retrieval} (CLEVER), that is able to incrementally index new documents while supporting the ability to query both newly encountered documents and previously learned documents. 
To this end, we need to resolve two key challenges in terms of the indexing and retrieval task.  

First, \emph{how to incrementally index new documents with low computational and memory costs?} We introduce \emph{incremental product quantization} (IPQ) based on product quantization (PQ) methods~\cite{jegou2010product} to generate PQ codes for documents as docids, which can represent large volumes of documents via a small number of quantization centroids. 
The key idea is to incrementally update a subset of centroids instead of all centroids, without the need to update the indices of existing data.   
Specifically, given the base documents (that is, the initial collection of documents), we iteratively train the document encoder and quantization centroids with a clustering loss and a contrastive loss. 
The clustering loss offers incentives for representations of documents around a centroid to be close, while the contrastive loss enhances the document representation close to its own random spans. 
This helps learn discriminative documents and centroid representations, so as to easily generalize to new documents. 
Then, as new documents arrive, we introduce two adaptive thresholds based on the distances between new and old documents in the representation space, to automatically realize three types of update for centroid representations, i.e., unchanging, changing, and addition.   
Finally, we index each new document by learning a mapping from document content to its docid. 

Second, \emph{how to prevent catastrophic forgetting for previously indexed documents and maintain the retrieval ability?} 
We take inspiration from the given-new strategy in cognitive science, in which humans attach new information to already known, i.e., given, similar information, in their memory to enhance a mental model of the information as a whole \citep{clark1974psychological, haviland1974s, clark1977discourse}. 
We propose a memory-augmented learning mechanism to strengthen connections between new and old documents. 
We first allocate a dynamic memory bank for each session to preserve exemplar documents similar to new documents to prevent forgetting of previously indexed documents.  
Then, we train a query generator model to sample pseudo-queries for documents and supplement them while continually indexing new documents to prevent forgetting for the retrieval task. 

\heading{Experimental findings}
We introduce two novel benchmark data\-sets constructed from the existing MS MARCO \cite{msmarco} and Natural Questions \cite{nq} datasets, simulating the continual addition of documents to the system. 
Extensive evaluation shows that CLEVER performs significantly better than prevailing continual learning methods and effectively mitigates catastrophic forgetting in incremental scenarios, while outperforming traditional IR models and existing GR models in non-incremental scenarios. 

\vspace*{-2mm}
\section{Problem Statement} 
\label{sec:formulation}


\textbf{Task formulation.}
Given a large-scale base document set $\mathcal{D}_0$ and sufficiently many labeled query-document pairs $\mathcal{P}_{0}^{\mathcal{D}_0}$, we can train an initial GR model $f(\cdot)$ via a standard seq2seq objective \cite{sutskever2014sequence}. 
Let the meta-parameters of the initial model be $\Theta_0$. 
The continual document learning task assumes the existence of $T$ new datasets $\{\mathcal{D}_1, \dots, \mathcal{D}_t, \dots, \mathcal{D}_T \}$, from $T$ sessions arriving in a sequential manner. 
In any session $t \ge 1$, $\mathcal{D}_t$ is only composed of newly encountered documents $\{d_t^1, d_t^2, \dots\}$ without queries related to these documents. 
Let the model parameters before the $t$-th update be $\Theta_{t-1}$. 
For session $t$, the GR model is trained to update its parameters to $\Theta_t$ via the new dataset $\mathcal{D}_t$ and previous datasets $\{\mathcal{D}_0, \dots, \mathcal{D}_{t-1}\}$, and $\Theta_t$ serves as input for the datasets $\{\mathcal{D}_0, \dots, \mathcal{D}_{t}\}$.   

\heading{Evaluation}
After updating \ac{GR} models with new documents, we explore two types of  test query set for performance evaluation.  

\begin{figure}[t]
 \centering
 \includegraphics[width=\columnwidth]{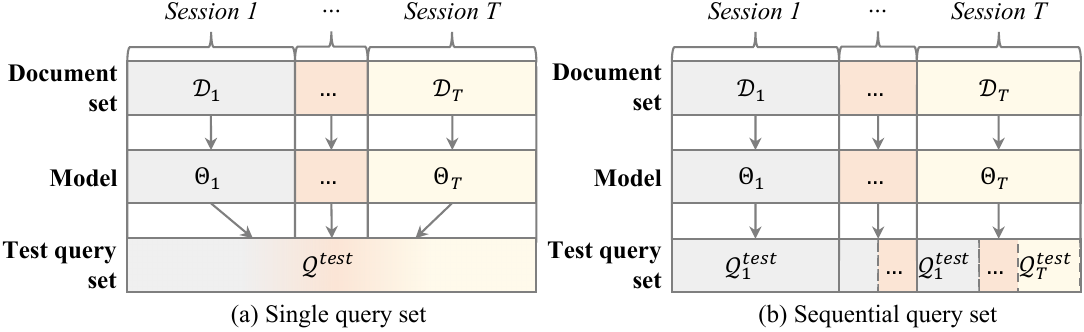}
 \caption{Evaluation criteria.}
 \label{fig:settings}
 \vspace*{-1mm}
\end{figure}

\textit{Single query set}. As illustrated in Figure~\ref{fig:settings} (a), under this condition, there is only one test query set $\mathcal{Q}^\mathit{test}$, and their relevant documents arrive in different sessions. 
However, we cannot directly compare the retrieval performance before and after incremental updates. 
The reason is that many widely-used ranking metrics \cite{manning2008introduction} are based on ground-truth relevant documents, which change across sessions. 
Instead, we compare the overall performance $\mathrm{VERT}_t$ of different methods on $\mathcal{Q}^\mathit{test}$ in the same session $t$ vertically,  
\begin{equation}
\mathrm{VERT}_t=\sum_{q\in \mathcal{Q}^\mathit{test}, d^+_q\in \{\mathcal{D}_0, \dots, \mathcal{D}_t\}}g(d^+_q, f(q; \Theta_t)),
\end{equation}
where $d^+_q$ is a relevant document to the query $q \in \mathcal{Q}^\mathit{test}$ in existing sessions $\{0, \dots, t\}$, and $g(\cdot)$ denotes a widely-used evaluation metric for IR; see Section~\ref{evaluation metirc}. 

\textit{Sequential query set}. 
As illustrated in Figure \ref{fig:settings} (b), under this condition, the test query set $\mathcal{Q}^\mathit{test}_t$ is specific for each session $t$, and the relevant documents appear in existing sessions $\{0, \dots, t\}$. 
We can directly compare different models across different sessions. 
Besides $\mathrm{VERT}$, following~\cite{gem,mehta2022dsi++}, we apply 
\begin{enumerate*}[label=(\roman*)]
\item average performance (AP) to measure the average performance by the end of training with the entire existing data sequence, 
\item backward transfer (BWT) to measure the influence of learning a new session on the preceding sessions' performance, and
\item forward transfer (FWT) to measure the ability to learn when presented with a new session. 
\end{enumerate*} 

\begin{figure*}[t]
 \centering
 \includegraphics[width=0.92\textwidth]{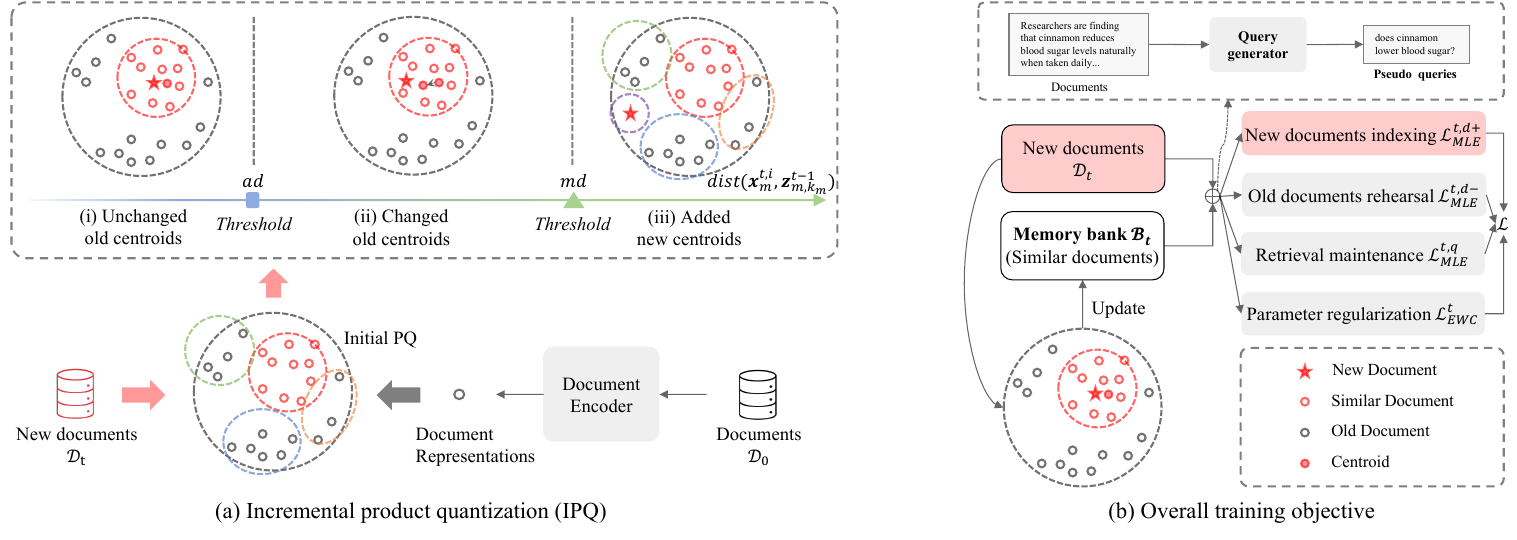}
 \caption{(a) Encoding new documents into docids by updating a subset of quantization centroids.  (b) Overall training objective for continual indexing while  alleviating forgetting of the retrieval ability.}
 \label{fig:overview}
 \vspace*{-4mm}
\end{figure*}

\vspace*{-2mm}
\section{Methodology}
In this section, we introduce our \emph{Continual-LEarner for generatiVE Retrieval} (CLEVER).  
Given an already constructed \ac{GR} model, we first index newly arrived documents (Section \ref{sec:indexing}), and then prevent forgetting of the retrieval ability during incremental indexing (Section \ref{DM}). 
Figure \ref{fig:overview} provides an overview of the method. 

\vspace*{-2mm}
\subsection{Indexing new documents} 
\label{sec:indexing}

To incrementally index new documents, we need to encode new documents into docids with low computational cost, while learning associations between new documents and their docids. 
\vspace*{-2mm}

\subsubsection{Incremental product quantization}
\label{sec:IPQ}
One popular docid representation is to leverage the \emph{product quantization} (PQ) technique \cite{jegou2010product} to generate the PQ code as the docid. 
PQ is able to produce a number of centroids with low storage costs, contributing to representing large collections of documents. 
However, it is not designed for dynamic corpora. 
Therefore, we propose \emph{incremental product quantization} (IPQ) based on PQ to represent docids. 

The key idea is to design two adaptive thresholds to update a subset of centroids instead of all centroids,  without changing the index of the updated centroids.  
IPQ contains two dependent steps: 
\begin{enumerate*}[label=(\roman*)]
\item construct the document encoder and base quantization centroids, from the base documents $\mathcal{D}_0$, and 
\item partially update quantization centroids, based on the relationship between new documents $\mathcal{D}_t$ and old documents $\{\mathcal{D}_0,\dots,\mathcal{D}_{t-1}\}$.  
\end{enumerate*}

\heading{Building base quantization centroids} Given the base document set $\mathcal{D}_0$, we first leverage BERT \cite{bert} as the initial document encoder.  
Specifically, a special token $w_0$ = \texttt{[CLS]} is added in front of the $i$-th document \smash{$d_0^{i}=\{w_1, \dots, w_{\vert d_0^{i} \vert}\}$} in $\mathcal{D}_0$, and the encoder represents the document $d_0^{i}$ as a series of hidden vectors, i.e., $\textbf{h}_0, \textbf{h}_1, \dots, \smash{\textbf{h}_{\vert d_0^{i} \vert}}=\operatorname{Encoder}(w_0, w_1, \dots, \smash{w_{\vert d_0^{i} \vert}})$. 
We feed the \texttt{[CLS]} representation $\textbf{h}_0$ into a projector network \cite{Chen2020ASF, Chen2021ExploringSS}, which is a feed-forward neural network with a non-linear activation function (i.e., $\tanh$), to obtain the complete document representation $\textbf{x}^{0,i}$ of $d_0^{i}$. 

To better generalize to new documents, we propose a two-step iterative process to iteratively learn document encoder and quantization centroids, to enhance their discriminative abilities. In Step 1, centroids are obtained via a clustering process over document representations, and in Step 2, document representations are learned from centroids with a bootstrapped training process.  

\textbf{Step 1: Clustering process for centroids.} Critically, given the $D$-dimensional document representation $\textbf{x}^{0, i} \in \mathbb{R}^D$ of the $i$-th document $d_0^{i} \in \mathcal{D}_0$, there are three main stages to build centroid representations (more generally the PQ codes) following \cite{jegou2010product}: 

\textbf{Group division}. We use $M$ sub-quantizers to divide the $D$-dimensional space into $M$ groups, i.e., $\textbf{x}^{0, i} \in \mathbb{R}^D$ is represented as a concatenation of $M$ sub-vectors $[\textbf{x}_1^{0, i} , \dots, \smash{\textbf{x}_m^{0, i}}, \dots, \smash{\textbf{x}_M^{0, i}} ]$, where $\smash{\textbf{x}_m^{0, i}} \in \mathbb{R}^{ D/M}$. In this way, the sub-vectors $\{\textbf{x}_m^{0, i}\}$ of each document $d_0^i \in \mathcal{D}_0$, where $i \in \{1, \dots, \vert \mathcal{D}_0 \vert\}$, form the $m$-th group.

\textbf{Group clustering}. For each group, we apply $K$-means clustering over all  document representations in $\mathcal{D}_0$ to obtain the initial codebook $Z^0 = \{ \smash{\textbf{z}_{m, k}^0}\}$, where $m \in \{ 1, \dots, M \}$, $k \in \{ 1, \dots, K \}$, and \smash{$\textbf{z}_{m, k}^0$} is the centroid of the $k$-th cluster from the $m$-th group at initial session. 
The codebook is composed of $M$ sub-codebooks, each of which contains $K$ cluster centroids quantized from a distinct sub-quantizer.

\textbf{Quantization}. Given the sub-vector $\smash{\textbf{x}_m^{0, i}\in \mathbb{R}^{D/M}}$, we quantize it to the nearest centroid \smash{$\textbf{z}_{m, k_m}^0$}. 
We select the centroid  $\textbf{z}_{m,\phi(\textbf{x}_m^{0, i})}^0$ which achieves the minimum quantization error and decide the $k_m$-th cluster in the $m$-th group that $\textbf{x}_m^{0, i}$ belongs to  
$
    k_m=\phi(\textbf{x}_m^{0, i}) = \arg\min_k\| \smash{\textbf{z}_{m, k}^0} - \textbf{x}_m^{0, i} \|^2. 
$
Finally, document representation $\textbf{x}^{0, i}$ is quantized as the concatenation of $M$ centroid representations, i.e., $\textbf{x}^{0, i} = [\smash{\textbf{z}_{1,k_1}^0}, \dots, \smash{\textbf{z}_{m,k_m}^0}, \dots, \smash{\textbf{z}_{M,k_M}^0} ]$.
Accordingly, docid $u_{d_0^i}$ of document $d_0^i$ is obtained by its PQ code $[k_1$, \dots, $k_m$, \dots, $k_M]$.

\textbf{Step 2: Bootstrapped training for document representations.} 
High-quality document representation is the foundation of PQ to support effective clustering. 
However, based on the original BERT, the discriminative ability of these representations may be limited since the representation will focus more on the common words and thus is not differential with other representations \cite{xu2022improving,ma2022pre}. 
Therefore, we propose a bootstrapped training process based on BERT to learn discriminative document representations. 
The key idea is to utilize both the contrastive loss and the clustering loss for re-training the BERT encoder itself.  

The \textbf{contrastive loss} helps to generate the document representation close to its own random spans while being far away from others \cite{ma2022pre}. We first sample a set of spans at four levels of granularity for each document with length $n$ in $\mathcal{D}_0$, including word-level, phrase-level, sentence-level and paragraph-level: 
\begin{enumerate*}[label=(\roman*)]
    \item \textit{length sampling}: We first sample the span length from a beta distribution for each level of granularity, i.e., $ \ell_{\mathit{span }}=p_{\mathit{span}} \cdot\left(\ell_{\mathit{max} }-\ell_{\mathit{min} }\right)+\ell_{\mathit{min} },$ where $\ell_{\mathit{min} }$ and $\ell_{\mathit{max} }$ denote the minimum and maximum span length of each level of granularity and $p_{\mathit{span }}$ is sampled by $p_{\mathit{span}} \sim \operatorname{Beta}(\alpha, \beta),$ where $\alpha$ and $\beta$ are two hyperparameters, and 
            
    \item \textit{position sampling}: We randomly sample the starting position $start \sim U\left(1, n-\ell_{\mathit{span }}\right)$ and the ending position $\mathit{end}=\mathit{start}+\ell_{\mathit{span }}$. In this way, the final span is denoted as $\mathit{span}=[w_\mathit{start}, \dots, w_\mathit{end-1}]$. 
    \end{enumerate*} 
Given a mini-batch of $N$ documents, we can obtain $N$ whole document  representations and their span representations and the contrastive loss is,
$
\mbox{}\hspace*{-1.5mm}
\mathcal{L}_{CL}\!=\!\sum_{i=1}^{|\mathcal{D}_0|}\!-\frac{1}{4 G}\!\sum_{s \in S(i)}\!\!\!\log\!\frac{\exp \left(\operatorname{sim}\left(\mathbf{s}_i, \mathbf{s}_s\right) / \tau\right)}{\sum_{j=1}^{N *(4 G+1)}\!\mathbbm{1}_{[i \neq j]} \exp \left(\operatorname{sim}\left(\mathbf{s}_i, \mathbf{s}_j\right) / \tau\right)},
\!\mbox{}
$
where $G$ is the number of spans sampled per granularity, $S(i)$ is the index set of spans from $d_0^i$ with size $4G$, $\operatorname{sim}(\cdot)$ is the dot-product function and $\tau$ is the temperature hyper-parameter, and $\textbf{s}_i$ is the span representation, which is computed via the average pooling operation over the output word representations given by the encoder, i.e., $\textbf{s}_i=\operatorname{AvgPooling}(\textbf{h}_\mathit{start}, \dots, \textbf{h}_\mathit{end-1}) $.

The \textbf{clustering loss} computes the mean square error (MSE) between the document representations before and after quantization, enabling to cluster document representations around the centroid representations. 
Concretely, the MSE loss $\mathcal{L}_{MSE}$ is,  
$
	\mathcal{L}_{MSE}=\sum_{i=1}^{\vert \mathcal{D}_0 \vert} \| \textbf{x}^{0, i} - \hat{\textbf{x}}^{0, i}\|^2. 
$
%
We then re-train the BERT encoder via $\mathcal{L}_{CL} + \mathcal{L}_{MSE}$, and also adopt the \texttt{[CLS]} representation given by the re-trained encoder as the document representation. 

\textbf{Repeating Step 1 and Step 2}. Step 1 and Step 2 are repeated iteratively for $v$ epochs. 
Finally, we obtain the initial quantization centroids to build the PQ codes $\smash{\mathcal{U}_0^{\mathcal{D}_0}}=\{u_{d_0^1}, \dots, \smash{u_{d_0^{\vert \mathcal{D}_0 \vert}}}\}$ for $\mathcal{D}_0$, and the $\operatorname{Encoder}(\cdot)$ learned on $\mathcal{D}_0$ is fixed for later sessions. 

\heading{Adaptively updating quantization centroids} 
With the arrival of new documents $\mathcal{D}_t$ during session $t$, we first utilize the learned $\operatorname{Encoder}(\cdot)$ to obtain the document representations. 
Based on the representations of new and old documents, a simple method is to re-cluster them to obtain the novel PQ codes as the docids. 
However, this may incur high computational cost in updating all clustering results and re-training the \ac{GR} model based on the updated docids.

Ideally, we have a way to balance the trade-off between update efficiency and quantization error. 
Here, we introduce a partial codebook update strategy for this purpose. 
Specifically, we design three types of update for centroid representations in each sub-codebook, contributing to efficiency in memory and computational load: 
\begin{itemize}[leftmargin=*,nosep]

\item \textbf{Unchanged old centroids}: It is pointless to update the centroid when the features in some groups of new documents have a trivial contribution in their corresponding centroid update. 

\item \textbf{Changed old centroids}: It is possible that some features of new documents have a vital contribution to a centroid update. 

\item \textbf{Added new centroids}: We should add new centroids when new documents are significantly different from all old documents. 
\end{itemize}

\noindent%
To achieve the above update, we first divide the representation vector $\textbf{x}^{t, i}$ of each document $d_t^i \in \mathcal{D}_t$, into $M$ sets of sub-vectors and add each sub-vector $\smash{\textbf{x}_m^{t, i}}$ to the corresponding $m$-th group. 
Then, for each sub-codebook, we compute the Euclidean distance \cite{danielsson1980euclidean} between the newly arrived sub-vector $\smash{\textbf{x}_m^{t,i}}$ and its nearest centroid $\smash{\textbf{z}_{m,k_m}^{t-1}}$ based on the last session $t-1$, i.e., $\operatorname{dist}(\textbf{x}^{t, i}_m, \textbf{z}_{m,k_m}^{t-1})$. 
Finally, we devise two adaptive thresholds, i.e., $ad$ and $md$, according to this distance, to achieve three types of update.  

For each cluster in a sub-codebook, $ad$ is the average distance between each document sub-vector and the quantization centroid,
\begin{equation}
	ad=\frac{1}{\vert C_{m,k}^{t-1} \vert}\sum_{j=1}^{\vert C_{m,k}^{t-1} \vert} \operatorname{dist}(\textbf{x}_m^{t, j}, \textbf{z}_{m,k_m}^{t-1}),
\end{equation}
where $\smash{C_{m,k}^{t-1}}$ is the set of document indices assigned to the centroid $\smash{\textbf{z}_{m,k}^{t-1}}$. 
And $md$ is the maximum distance between each document sub-vector and the quantization centroid, denoted as,
\begin{equation}
	md=\max_{j\in C_{m,k}^{t-1}} \operatorname{dist}(\textbf{x}_m^{t, j}, \textbf{z}_{m,k_m}^{t-1}) + \operatorname{rand\_dist},
\end{equation}
where $\operatorname{rand\_dist} \sim U(0, ad)$ is sampled from the continuous uniform distribution. 
Note that the condition $ad \leq md$ always holds.

Therefore, as depicted in Figure \ref{fig:overview}(a), we can automatically decide the update type of each centroid representation as follows: 
\begin{itemize}[leftmargin=*,nosep]
	\item If $\smash{\operatorname{dist}(\textbf{x}^{t,i}_m, \textbf{z}_{m,k_m}^{t-1})} < ad$, the centroid  remains unchanged. 

    \item Alternatively, if $\smash{ad \leq \operatorname{dist}(\textbf{x}^{t,i}_m, \textbf{z}_{m,k_m}^{t-1})} \leq md$, we need to update the centroid representation. 
    We first update the set via $\smash{C_{m,k}^t}=\smash{C_{m,k}^{t-1}} \cup \{ \mathit{ind} \}$, where $\mathit{ind}$ is the index number of $\textbf{x}^{t,i}_m$. Then each centroid can be updated by,
    $
    \textbf{z}_{m,k_m}^{t} = \textbf{z}_{m,k_m}^{t-1} + \frac{1}{\vert C_{m,k}^t \vert}(\textbf{x}_m^{t,i}-\textbf{z}_{m,k_m}^{t-1}).
    $
  
  \item  Finally, if $\operatorname{dist}(\textbf{x}^{t,i}_m, \textbf{z}_{m,k_m}^{t-1}) > md$, we add a new cluster and thus there are $K+1$ clusters in the group.  We directly use the document sub-vector as the centroid representation, i.e., $ \textbf{z}_{m, K+1}^{t}=\textbf{x}_{m}^{t,i}$. 
\end{itemize}

\noindent%
After applying the above update strategy for all $M$ sub-codebooks, we obtain the specific codebook $Z^t$ at session $t$: 
\begin{enumerate*}[label=(\roman*)]
\item for new documents in $\mathcal{D}_t$, we obtain their PQ codes  $\mathcal{U}_t^{\mathcal{D}_t}$ based on the $Z^t$ as the docids, and
\item for old documents, their PQ codes will not be affected since we only operate on the centroid representations, instead of the index of the updated centroid. 
In the case of old documents around a centroid sharing the same representation, i.e., $ad=md=0$, we directly change the centroid representation based on the new document sub-vector. 
\end{enumerate*}

\vspace*{-2mm}
\subsubsection{Indexing objective}
To memorize information about each new document, we leverage maximum likelihood estimation (MLE) \cite{mle} to maximize the likelihood of a docid conditioned on the corresponding document, i.e., 
\begin{equation}
    \mathcal{L}_{MLE}^{t, d+} = \sum_{\mathcal{D}_t, \mathcal{U}_t^{\mathcal{D}_t} } \log p(u_{d_t^i}, \Theta_t \mid d_t^i; \Theta_{t-1}), 
\end{equation}
where $d_t^i\in \mathcal{D}_t$, $u_{d_t^i} \in \mathcal{U}_t^{\mathcal{D}_t}$, $\Theta_t$ is the \ac{GR} model parameters at the session $t$, and $i\in \{1, \dots, \vert \mathcal{D}_t \vert \}$.

\vspace*{-2mm}
\subsection{Preserving retrieval ability}
\label{DM}

During continual indexing of new documents, it is important for the \ac{GR} models to prevent forgetting the retrieval ability.  
We are inspired by the fact that humans benefit from previous similar experiences when taking actions \cite{clark1974psychological, haviland1974s, clark1977discourse} and propose a memory-augmented learning mechanism to build meaningful connections between new and old documents. 
Specifically, we first construct a memory bank with similar documents for each new session and replay  the process of indexing them alongside the indexing of new documents. 
Then, we leverage a query generator model to sample pseudo-queries for documents and the resulting query-docid pairs are employed to maintain the retrieval ability.   
The overall learning process is visualized in Figure~\ref{fig:overview} (b). 

\heading{Dynamic memory bank construction} 
The memory bank is allocated to store a tiny subset of old documents which are similar to new documents in the PQ space. 
 We assume that two documents are similar if many dimensions of their PQ codes are the same.  
For each document in $\mathcal{D}_t$, we target to retrieve its similar documents at different levels. 
Concretely, we iteratively change its PQ code at different dimensions, which includes the following steps: 
 \begin{enumerate*}[label=(\roman*)]
    \item we first set the number $o$ of PQ code dimensions that will be changed to $1$; 
    \item we randomly select $o$ dimensions of the PQ code and  assign different centroids to the selected dimensions to obtain the similar PQ code. We repeat this process $c$ times; and 
    \item we obtain the similar documents from the previous sessions  if they are associated with the obtained PQ codes. 
\end{enumerate*}
The processes in (ii) and (iii) are repeated by increasing $o$ with 1 to at most ${M}/{6}$. 
 
Finally, we group the similar documents of each document in  $\mathcal{D}_t$ to construct a specific memory bank $\mathcal{B}_t$ at the session $t$. 
Note that the memory bank is dynamically updated at each new session.

\heading{Rehearsing the indexing of old documents} 
For each new session $t$, we aim to prevent forgetting previously indexed documents.
Given the meta model parameters $\Theta_{t-1}$ before the $t$-th update, we apply MLE over the memory bank $\mathcal{B}_t$ to update the \ac{GR} model, i.e., 
\begin{equation}
    \mathcal{L}_{MLE}^{t, d-} = \sum_{d_t^i\in \mathcal{B}_t,  u_{d_t^i} \in  \mathcal{U}_t^{\mathcal{B}_t} } \log p(u_{d_t^i}, \Theta_t \mid d_t^i; \Theta_{t-1}), 
\end{equation}
where $\mathcal{U}_{t}^{\mathcal{B}_t}$ is the PQ codes of $\mathcal{B}_t$, and $i\in \{1, \dots, \vert \mathcal{B}_t \vert \}$.

\heading{Constructing pseudo query-docid pairs}
To prevent forgetting the retrieval ability during indexing new documents, we train a query generator model to sample pseudo-queries for documents and supplement the query-docid pairs during indexing.    
We fine-tune the T5 model \cite{t5} based on the query-document pairs $\mathcal{P}_{0}^{\mathcal{D}_0}$ in the initial session, by taking the document terms as input and producing a query following \cite{doc2query}. 
After fine-tuning, the model parameters of the query generator $\Theta_{qg}$ are fixed.   

For each new session $t$, we generate pseudo-queries for each document in $\mathcal{D}_t$ and $\mathcal{B}_t$ via $\Theta_{qg}$ and denote the obtained pairs of pseudo-queries and documents as $\mathcal{P}_t^{\mathcal{D}_t}$ and $\mathcal{P}_t^{\mathcal{B}_t}$, respectively. 
Given the meta model parameters $\Theta_{t-1}$ of the \ac{GR} model, we also apply MLE to maximize the likelihood of a relevant docid conditioned on each pseudo query in $\mathcal{P}_t^{\mathcal{D}_t}$ and $\mathcal{P}_t^{\mathcal{B}_t}$, denoted as,
\begin{equation}
    \mathcal{L}_{MLE}^{t, q} = \sum_{\{\mathcal{P}_t^{\mathcal{D}_t}, \mathcal{P}_t^{\mathcal{B}_t}\}, \{\mathcal{U}_t^{\mathcal{D}_t}, \mathcal{U}_t^{\mathcal{B}_t}\}} \log p(u_{d_t^{\psi(q_j)}}, \Theta_t \mid q_j;\Theta_{t-1}),
\end{equation}
where $(q_j, \smash{d_t^{\psi(q_j)}})\in \{\smash{\mathcal{P}_t^{\mathcal{D}_t}}, \smash{\mathcal{P}_t^{\mathcal{B}_t}}\}$ , $\psi(q_j)$ is the index of the relevant document to $q_j$ and $j\in \{1, \dots, \vert \{\smash{\mathcal{P}_t^{\mathcal{D}_t}}, \smash{\mathcal{P}_t^{\mathcal{B}_t}}\} \vert\}$.  
$\smash{u_{d_t^{\psi(q_j)}}}\in \{\smash{\mathcal{U}_t^{\mathcal{D}_t}},  \smash{\mathcal{U}_t^{\mathcal{B}_t}}\}$ is the relevant docid. 

\vspace*{-2mm}
\subsection{Overall training objective}

In the training phase, we sequentially train the \ac{GR} model on each session $t$ by combining the objective for indexing and retrieval, i.e., 
\begin{equation} \label{eq:objective}
    \min_{\Theta_t} -(\mathcal{L}_{MLE}^{t, d+} + \mathcal{L}_{MLE}^{t, d-} + \mathcal{L}_{MLE}^{t, q})+\lambda \mathcal{L}_{EWC}^t, 
\end{equation}
where $\lambda$ is a hyper-parameter. The elastic weight consolidation (EWC) \cite{ewc} loss $\mathcal{L}_{EWC}^t$ is used to regularize the model parameters, via the weighted distance between  $\Theta_{t-1}$ and $\Theta_t$,  
\begin{equation}
      \mathcal{L}_{EWC}^t=\sum_l F_l\left(\Theta_{t-1, l}-\Theta_{t, l}\right)^2,
\end{equation}
where $F$ is the Fisher information matrix \cite{ewc}, and $F_l$ denotes each model parameter.

\vspace*{-1mm}
\section{Experimental Settings}

\vspace*{-1mm}
Next, we summarize our experimental settings. The code can be found at \url{https://github.com/ict-bigdatalab/CLEVER}.

\vspace*{-2mm}
\subsection{Benchmark construction}

To facilitate the study of continual document learning for \ac{GR}, we build two benchmark datasets, i.e., CDI-MS and CDI-NQ, from MS MARCO 300k \cite{msmarco, mehta2022dsi++, zhou2022ultron} and Natural Questions (NQ) \cite{nq}, respectively. 
MS MARCO 300k is a subset of the MS MARCO Document Ranking collection, comprising 300k documents.
NQ contains 307k query-document training pairs, 231k documents, and 7.8k queries in the dev set. 
We report the performance results on the dev sets as both MS MARCO 300k and NQ leaderboard limit the frequency of submissions~\citep{dsi, mehta2022dsi++}. 

To mimic the new arrival of documents in MS MARCO and NQ, we first randomly sample $60\%$ documents from the whole document set as the base documents $\mathcal{D}_0$, and 
    leverage their corresponding relevance labels to construct the query-document pairs $\mathcal{P}_{0}^{\mathcal{D}_0}$. 
Then, we randomly sample $10\%$ documents from the remaining document set as the new document set, and this operation is repeated for 4 times to obtain $\mathcal{D}_1$--$\mathcal{D}_4$. 
The test query set is defined as follows:
\begin{enumerate*}[label=(\roman*)]
    \item for a single query set, all dev queries are denoted as $\mathcal{Q}^\mathit{test}$, and 
    \item for sequential query set, we sample $60\%, 10\%, 10\%, 10\%$ and $10\%$ queries from the whole dev query set as $\mathcal{Q}^\mathit{test}_0$, \ldots, $\mathcal{Q}^\mathit{test}_4$, respectively.
\end{enumerate*}

\vspace*{-2mm}
\subsection{Models}

\vspace*{-1mm}
\subsubsection{Baselines}
\heading{Traditional IR models} 
\begin{enumerate*}[label=(\roman*)]
	\item \textbf{BM25} \cite{bm25} is an effective sparse retrieval model and we re-index all previously seen documents upon the arrival of new documents.
	\item \textbf{DPR} \cite{dpr} is a representative dense retrieval model with BERT-based dual-encoder architecture. We use the model trained on the first session $\mathcal{D}_0$ to encode newly arrived documents and then add their encoded representations to the existing external index. 
\end{enumerate*}

\heading{Generative retrieval models}
\begin{enumerate*}[label=(\roman*)]
	\item \textbf{DSI} \cite{dsi} encodes  all information about the corpus in the model parameters and we adopt atomic docids in DSI.  
	\item \textbf{DSI-QG} \cite{zhuang2022bridging} leverages a query generation model to augment the document collection at indexing. 
	\item \textbf{NCI} \cite{nci} utilizes a prefix-aware weight-adaptive decoder and we adopt NCI with DocT5Query for augmented queries.  
	\item \textbf{Ultron} \cite{zhou2022ultron} applies a three-stage training pipeline and we adopt Ultron with PQ as the docid.
\end{enumerate*}	
Due to their limitations in accommodating dynamic corpora, we only evaluate the performance in non-incremental scenarios. 

Furthermore, we compare our model with an adaption of Ultron as the \textbf{BASE} method, wherein PQ technique is used to generate docids and the GR model is continually fine-tuned by directly mapping each new document to its docid. 
We also compare with \textbf{DSI++} \cite{mehta2022dsi++}, which continuously fine-tunes DSI over new documents by directly assigning each new document an atomic docid, i.e., an arbitrary unique integer. We re-implement it since the source code has not yet been released.  

\vspace*{-2mm}

\subsubsection{Model variants} 

To verify the effectiveness of IPQ, we implement variants with the memory-augmented learning mechanism. 
To build base quantization centroids, we have  
\begin{enumerate*}[label=(\roman*)]
    \item \textbf{CLEVER$_\mathit{atomic}$}, which uses arbitrary unique integers as docids, as used in DSI++;
    \item \textbf{CLEVER$_{PQ}$}, which directly uses the original BERT$_\mathit{base}$ \cite{bert} to obtain document representations, and builds PQ codes as docids by the original PQ technique \cite{jegou2010product}; the codebook is fixed in all sessions; and 
    \item \textbf{CLEVER$_{PQ+Re}$}, which extends CLEVER$_{PQ}$ by re-clustering the document representations obtained by BERT$_{base}$ as new documents arrive; the codebook is updated at each new session. 
\end{enumerate*} 
To adaptively update the quantization centroids, the variants are: 
\begin{enumerate*}[label=(\roman*)]
	\item \textbf{CLEVER$_{PQ+Dis}$} leverages the two-step iterative process to build discriminative base PQ codes; then, the codebook is fixed for new sessions, i.e., only adopting the ``unchanged old centroids'' type;  
    \item \textbf{CLEVER$_{PQ+Dis+ad}$} extends \textbf{CLEVER$_{PQ+Dis}$} by adding the $ad$ threshold, i.e., adopting  ``unchanged old centroids'' and ``changed old centroids;'' 
    \item \textbf{CLEVER$_{PQ+Dis+md}$} extends \textbf{CLEVER$_{PQ+Dis}$} by adding the $md$ threshold, i.e., adopting ``added new centroids'' and ``changed old centroids'' to update the quantization centroids; and
     \item  \textbf{CLEVER$_{PQ+Dis+Re}$} extends \textbf{CLEVER$_{PQ+Dis}$} by re-building discriminative PQ codes for all documents as new documents arrive; the codebook is updated at each new session. 
\end{enumerate*}

To verify the effectiveness of the memory-augmented learning mechanism, variants (while using IPQ) are: 
\begin{enumerate*}[label=(\roman*)]
    \item \textbf{CLEVER$^{-EWC}$}, which removes $\mathcal{L}_{EWC}^t$ in Eq.~\ref{eq:objective} to re-train the \ac{GR} model;

    \item \textbf{CLE\-VER$^{-MLE(d-)}$}, which removes $\mathcal{L}_{MLE}^{t, d-}$ in Eq.~\ref{eq:objective} to re-train the \ac{GR} model;

    \item \textbf{CLEVER$^{-MLE(q)}$}, which removes $\mathcal{L}_{MLE}^{t, q}$ in Eq.~\ref{eq:objective} to re-train the \ac{GR} model; and
        
    \item \textbf{CLEVER$^{Random}$}, which randomly selects some old documents to construct the memory bank with the same number of similar documents in CLEVER, which is an adaption of DSI++ \cite{mehta2022dsi++}, and then re-trains the \ac{GR} model via Eq.~\ref{eq:objective}. 
\end{enumerate*}

\vspace*{-2mm}
\subsection{Evaluation metrics}
\label{evaluation metirc}

The evaluation metric $g(\cdot)$ for IR in Section \ref{sec:formulation} is usually taken to be mean reciprocal rank (MRR@$N$), recall (R@$N$), hit ratio (Hits@$N$) and top-$N$ retrieval accuracy (ACC$@N$).   
Following~\cite{dsi, mehta2022dsi++, nci, zhou2022ultron}, we  show the continual results in terms of MRR$@10$ and HIT$@10$ for CDI-MS and CDI-NQ, respectively. 
By conducting further analyses, we find that the relative order of different models on other IR metrics is quite consistent with that on the MRR$@10$ and Hits$@10$.

\vspace{-2mm}
\subsection{Implementation details}
\label{sec:implement}

For IPQ, the length $M$ of PQ codes is 24, the number of clusters $K$ is 256, and the dimension of vectors $D$ is 768.
For the contrastive loss in the document encoder, $\ell_{\mathit{min} }$ and $\ell_{\mathit{max} }$ for sampling phrase-level spans are 4 and 16, respectively.
For sentence-level spans, $\ell_{\mathit{min} }$ and $\ell_{\mathit{max} }$ are  16 and 64, respectively.
For paragraph-level spans, $\ell_{\mathit{min} }$ and $\ell_{\mathit{max} }$ are  64 and 128, respectively.
The $\alpha$ and $\beta$ in the beta distribution are 4 and 2, respectively, which skews sampling towards longer spans. 
The number of spans sampled per granularity $G$ is 5. 
For the memory-augmented learning mechanism, the repeat time $c$ is 10, the probability $p$ is 0.2, the scale $r$ is 0.2, and $\lambda$ is 0.5. 

To train the document encoder in IPQ, we initialize the document encoder from the official BERT's checkpoint.
We use a learning rate of $5e^{-5}$ and Adam optimizer \cite{adam} with a linear warmup over the first 10\% steps.
Long input documents are truncated into several chunks with a maximum length of 512. 
The hyper-parameter of $\tau$ is 0.1.
We train for 6 epochs on four NVIDIA Tesla A100 40GB GPUs.

The \ac{GR} baselines and all variants of CLEVER, are based on the transformer-based encoder-decoder architecture, where the hidden size is 768, the feed-forward layer size is 3072, the number of transformer layers is 12, and the number of self-attention heads is 12, for both encoder and decoder.  
We implement the generative model in PyTorch based on Huggingface's Transformers library.
We initialize the parameters of the encoder-decoder architecture from the official checkpoint of T5$_{base}$ ~\cite{t5}. 
We use a learning rate of $3e^{-5}$ and Adam optimizer with the warmup technique, where the learning rate increases over the first 10\% of batches, and then decays linearly to zero.
The max length of the input is 512, the label smoothing is 0.1, the weight decay is 0.01, and the gradient norm clipping is 0.1.
We train in batches of 8192 tokens on four NVIDIA Tesla A100 40GB GPUs. 
At inference time, we adopt constrained beam search \cite{genre} to decode the docids with 24 timesteps and 15 beams. 
To train the query generator, we also initialize the parameters from the official checkpoint of T5$_{base}$ ~\cite{t5}, with a learning rate of $5e^{-4}$.  
For each new document, we adopt beam search to decode the pseudo queries with utmost 32 timesteps and 10 beams.

\vspace*{-3mm}
\section{Experimental Results}

\begin{table*}[t]
\small
    \caption{Model performance under the \textit{single query set} setting. We evaluate the  performance of $\mathcal{Q}^\mathit{test}$ in each session $\mathcal{D}_0$--$\mathcal{D}_4$ in terms of VERT (\%), respectively. 
    $*$ indicates statistically significant improvements over all baselines (p-value $< 0.05$).}
    \label{tab:single-res}
    \centering
    \setlength{\tabcolsep}{11pt}
    \renewcommand{\arraystretch}{0.75}
    \begin{tabular}{l ccccc  ccccc}
        \toprule
         & \multicolumn{5}{c}{\textbf{CDI-MS (MRR@10)}} &  \multicolumn{5}{c}{\textbf{CDI-NQ (Hits@10)}} \\
        \cmidrule(r){2-6} \cmidrule{7-11}
        \textbf{Model} & \multicolumn{1}{c}{\textbf{$\mathcal{D}_0$}} & \multicolumn{1}{c}{\textbf{$\mathcal{D}_1$}} & \multicolumn{1}{c}{\textbf{$\mathcal{D}_2$}} & \multicolumn{1}{c}{\textbf{$\mathcal{D}_3$}} & \multicolumn{1}{c}{\textbf{$\mathcal{D}_4$}} & \multicolumn{1}{c}{\textbf{$\mathcal{D}_0$}} & \multicolumn{1}{c}{\textbf{$\mathcal{D}_1$}} & \multicolumn{1}{c}{\textbf{$\mathcal{D}_2$}} & \multicolumn{1}{c}{\textbf{$\mathcal{D}_3$}} & \multicolumn{1}{c}{\textbf{$\mathcal{D}_4$}} \\
        \midrule
           BM25 & 28.47 & 29.12 & 29.33 & 29.24 & 28.96 & 35.65 & 35.46 & 35.16 & 36.88 & 35.23 \\
          DPR & 38.51 & 34.25 & 28.07 & 26.34 & 23.68 & 40.37 & 36.92 & 31.10 & 28.64 & 26.78 \\
		BASE & 42.78 & 39.51 & 38.66 & 36.16 & 34.93 & 67.13 & 64.45 & 62.08 & 59.24 & 58.61 \\
        DSI++ & 42.54 & 41.60 & 41.12 & 39.58 & 38.29 & 66.05 & 65.66 & 64.47 & 63.51 & 63.26 \\
		\midrule
		CLEVER$_\mathit{atomic}$ & 42.54 & 42.52 & 41.73 & 41.80 & 41.21 & 66.05 & 66.37 & 66.58 & 65.82 & 65.41 \\
		CLEVER$_{PQ}$ & 42.78 & 41.53 & 40.97 & 39.10 & 38.04 & 67.13 & 66.09 & 64.39 & 64.26 & 63.57 \\
	    CLEVER$_{PQ+Re}$ & 42.78 & 40.39 & 39.15 & 37.18 & 36.48 & 67.13 & 65.30 & 64.15 & 63.26 & 62.33 \\
		CLEVER$_{PQ+Dis}$ & 44.96 & 42.58 & 41.60 & 40.26 & 39.50 & 68.74 & 67.02 & 65.47 & 64.95 & 64.22 \\
		CLEVER$_{PQ+Dis+ad}$ & 44.96 & 44.03 & 43.78 & 42.81 & 42.06 & 68.74 & 67.43 & 67.69 & 66.30 & 66.15 \\
		CLEVER$_{PQ+Dis+md}$ & 44.96 & 43.81 & 43.59 & 42.30 & 41.74 & 68.74 & 66.85 & 66.34 & 65.92 & 65.28\\
		CLEVER$_{PQ+Dis+Re}$ & 44.96 & 42.27 & 40.53 & 38.64 & 37.90 & 68.74 & 65.60 & 64.28 & 63.79 & 62.83\\
		\midrule
		CLEVER$^{-EWC}$ & 44.96 & 43.51 & 42.67 & 41.27 & 41.30 & 68.74 & 67.16 & 66.84 & 66.75 & 66.09 \\
        CLEVER$^{-MLE(d-)}$ & 44.96 & 42.55 & 42.01 & 40.87 & 40.33 & 68.74 & 66.81 & 66.07 & 65.39 & 64.83 \\
        CLEVER$^{-MLE(q)}$ & 44.96 & 42.07 & 41.51 & 39.52 & 39.47 & 68.74 & 66.02 & 66.11 & 64.94 & 64.28 \\
		CLEVER$^\mathit{Random}$ & 44.96 & 41.83 & 41.24 & 39.76 & 38.62 & 68.74 & 65.37 & 64.24 & 63.81 & 63.10 \\
		\midrule
		CLEVER & \textbf{44.96}\rlap{$^*$} & \textbf{45.36}\rlap{$^*$} & \textbf{44.81}\rlap{$^*$} & \textbf{44.07}\rlap{$^*$} & \textbf{43.75}\rlap{$^*$} & \textbf{68.74}\rlap{$^*$} & \textbf{68.25}\rlap{$^*$} & \textbf{68.36}\rlap{$^*$} & \textbf{67.71}\rlap{$^*$} & \textbf{67.50}\rlap{$^*$} \\        
        \bottomrule
    \end{tabular}
    \vspace*{-2mm}
\end{table*}

\begin{table*}[t]
\small
    \caption{Model performance under the \textit{sequential query set} setting.  We evaluate the performance of $\mathcal{Q}^\mathit{test}_0$, \ldots, $\mathcal{Q}^\mathit{test}_4$ in each session $\mathcal{D}_0$--$\mathcal{D}_4$ in terms of VERT (\%), respectively. For AP, BWT and FWT,  $\uparrow$ indicates higher is better, and $\downarrow$ indicates lower is better. $*$ indicates statistically significant improvements over all baselines (p-value $< 0.05$).}
    \label{tab:seq-res}
    \centering
    \setlength{\tabcolsep}{3.4pt}
    \renewcommand{\arraystretch}{0.75}
    \begin{tabular}{l@{ \ } ccccc ccc ccccc ccc}
        \toprule
        & \multicolumn{8}{c}{\textbf{CDI-MS (MRR@10)}} &  \multicolumn{8}{c}{\textbf{CDI-NQ (Hits@10)}} \\
        \cmidrule(r){2-9} \cmidrule{10-17}
        \textbf{Model} & \multicolumn{1}{c}{\textbf{$\mathcal{D}_0$}} & \multicolumn{1}{c}{\textbf{$\mathcal{D}_1$}} & \multicolumn{1}{c}{\textbf{$\mathcal{D}_2$}} & \multicolumn{1}{c}{\textbf{$\mathcal{D}_3$}} & \multicolumn{1}{c}{\textbf{$\mathcal{D}_4$}} & \textbf{AP$\uparrow$} & \textbf{BWT$\downarrow$} & \textbf{FWT$\uparrow$} & \multicolumn{1}{c}{\textbf{$\mathcal{D}_0$}} & \multicolumn{1}{c}{\textbf{$\mathcal{D}_1$}} & \multicolumn{1}{c}{\textbf{$\mathcal{D}_2$}} & \multicolumn{1}{c}{\textbf{$\mathcal{D}_3$}} & \multicolumn{1}{c}{\textbf{$\mathcal{D}_4$}} & \textbf{AP$\uparrow$} & \textbf{BWT$\downarrow$} & \textbf{FWT$\uparrow$} \\
        \midrule
        BM25 & 29.63 & 29.54 & 28.30 & 28.12 & 28.37 & 27.99 & 1.00 & 28.58 & 34.93 & 33.67 & 33.58 & 32.83 & 33.64 & 32.60 & 1.42 & 33.43 \\
        DPR & 40.48 & 35.61 & 33.73 & 30.07 & 26.75 & 31.54 & 2.24 & 31.54 & 39.56 & 37.24 & 35.93 & 32.96 & 30.71 & 33.03 & 2.81 & 34.21 \\
   BASE & 43.64 & 40.08 & 37.91 & 36.55 & 35.27 & 32.91 & 7.23 & 37.45 & 68.43 & 65.38 & 63.19 & 60.57 & 59.02 & 56.55 & 8.46 & 62.04 \\
    DSI++ & 43.25 & 41.39 & 40.58 & 39.24 & 38.31 & 38.31 & 2.74 & 39.81 & 68.21 & 66.83 & 64.75 & 62.30 & 61.47 & 62.30 & 3.02 & 63.84 \\
		\midrule
		CLEVER$_\mathit{atomic}$ & 43.25 & 42.51 & 42.49 & 41.67 & 40.74 & 40.05 & 2.61 & 41.85 & 68.21 & 67.54 & 67.03 & 66.84 & 66.40 & 65.60 & 2.00 & 66.95 \\
		CLEVER$_{PQ}$ & 43.64 & 42.23 & 41.37 & 40.81 & 39.44 & 38.29 & 4.01 & 40.96 & 68.43 & 65.78 & 64.33 & 63.50 & 63.44 & 61.97 & 3.91 & 64.21 \\
		CLEVER$_{PQ+Re}$ & 43.64 & 41.92 & 41.03 & 38.41 & 37.16 & 36.95 & 4.36 & 39.63 & 68.43 & 65.28 & 63.05 & 62.83 & 62.41 & 61.02 & 4.22 & 63.39 \\
		CLEVER$_{PQ+Dis}$ & 45.30 & 43.26 & 42.70 & 41.57 & 40.21 & 39.47 & 3.93 & 41.93 & 69.57 & 66.54 & 65.84 & 64.01 & 64.27 & 62.23 & 3.53 & 65.17 \\
		CLEVER$_{PQ+Dis+ad}$ & 45.30 & 44.83 & 44.65 & 43.27 & 43.19 & 43.45 & 0.99 & 43.98 & 69.57 & 68.48 & 67.50 & 67.83 & 66.99 & 66.97 & 1.38 & 67.70 \\
	CLEVER$_{PQ+Dis+md}$ & 45.30 & 44.07 & 43.27 & 43.35 & 42.62 & 42.67 & 1.32 & 43.33 & 69.57 & 68.02 & 67.34 & 66.04 & 65.85 & 66.01 & 1.69 & 66.81 \\
		CLEVER$_{PQ+Dis+Re}$ & 45.30 & 42.83 & 41.04 & 39.59 & 38.77 & 38.26 & 4.06 & 40.56 & 69.57 & 67.31 & 65.48 & 64.02 & 63.25 & 63.02 & 3.64 & 65.02 \\
		\midrule
		CLEVER$^{-EWC}$ & 45.30 & 43.71 & 43.56 & 42.98 & 42.35 & 42.45 & 1.41 & 43.15 & 69.57 & 67.01 & 68.35 & 67.24 & 66.90 & 66.43 & 1.73 & 67.38 \\
		CLEVER$^{-MLE(d-)}$ & 45.30 & 43.40 & 42.76 & 42.03 & 42.11 & 41.58 & 1.92 & 42.58 & 69.57 & 68.74 & 67.52 & 66.36 & 65.38 & 65.10 & 3.02 & 67.00 \\
		CLEVER$^{-MLE(q)}$ & 45.30 & 42.81 & 42.09 & 41.74 & 41.88 & 40.80 & 2.45 & 42.13 & 69.57 & 66.32 & 65.61 & 65.06 & 64.40 & 63.56 & 3.29 & 65.35 \\
		CLEVER$^\mathit{Random}$ & 45.30 & 42.22 & 41.53 & 40.71 & 39.50 & 39.74 & 2.69 & 40.99 & 69.57 & 65.01 & 63.48 & 62.26 & 61.79 & 61.43 & 3.75 & 63.14 \\
		\midrule
		CLEVER & \textbf{45.30}\rlap{$^*$} & \textbf{45.26}\rlap{$^*$} & \textbf{45.09}\rlap{$^*$} & \textbf{44.85}\rlap{$^*$} & \textbf{44.71}\rlap{$^*$} & \textbf{44.39}\rlap{$^*$} & \textbf{0.82}\rlap{$^*$} & \textbf{44.98}\rlap{$^*$} & \textbf{69.57}\rlap{$^*$} &  \textbf{69.04}\rlap{$^*$} & \textbf{69.21}\rlap{$^*$} & \textbf{69.36}\rlap{$^*$} & \textbf{68.75}\rlap{$^*$} &  \textbf{68.56}\rlap{$^*$} & \textbf{0.78}\rlap{$^*$} & \textbf{69.09}\rlap{$^*$} \\ 
        \bottomrule
    \end{tabular}
    \vspace*{-3mm}
\end{table*}

In this section, we 
\begin{enumerate*}[label=(\roman*)]
\item analyze the retrieval performance on the CID-MS and CID-NQ datasets under both \textit{incremental} and \textit{non-incremental} settings, 
\item assess catastrophic forgetting and forward transfer abilities, and 
\item analyze the memory and computation cost. 
\end{enumerate*}
For (ii) and (iii), we conduct experiments on the CID-MS dataset under \textit{sequential query set} setting in terms of VERT(\%).

\vspace*{-4mm}
\subsection{Baseline comparison}

\textbf{Incremental performance on a single query set.} The performance comparisons between  CLEVER and  baselines on the \textit{single query set} are shown in Table~\ref{tab:single-res}. 
BM25 exhibits better performance than DPR and the underlying reason may be that BM25 is a data-independent probabilistic model, which renders it adaptable in the face of dynamic corpora.  
And the BASE method suffers a significant drop as new documents are added. By assigning new documents with atomic docids and using sampled old documents, DSI++ shows slight improvements over BASE.  
These results show that continual document learning for GR is a non-trivial challenge.

When we look at variants of CLEVER in terms of IPQ, we find that: 
\begin{enumerate*}[label=(\roman*)] 
\item CLEVER$_{PQ}$ performs worse than CLEVER$_{atomic}$ which updates the embeddings for each individual docid (also used in DSI++), showing that it is difficult for the \ac{GR} models to accommodate to new documents without updating docids. 
However, as shown in Section \ref{sec:cost}, for CLEVER$_{atomic}$, its memory continues to grow with the increase of documents and the time consumption of each update step increases with the number of steps. 
\item CLEVER$_{PQ+Re}$ and CLEVER$_{PQ+Dis+Re}$ give the worst performance. 
The reason might be that re-clustering old and new documents changes the previously learned centroids and thus the docids of old documents, which makes the learned document-docid mapping  invalid. 
\item The improvements of CLEVER$_{PQ+Dis}$ over CLEVER$_{PQ}$ demonstrate the need to learn discriminative document representations and quantization centroids. 
\item CLEVER$_{PQ+Dis+ad}$ and CLEVER$_{PQ+Dis+md}$ outperform CLEVER$_{PQ+Dis}$, showing that incorporating updates to old centroids and introducing new centroids could facilitate the assimilation of new documents. 
\end{enumerate*}

When we look at variants of CLEVER with different learning mechanisms, we observe that:
\begin{enumerate*}[label=(\roman*)]
\item CLEVER$^{-MLE(q)}$ performs worse than CLEVER$^{-MLE(d-)}$ and CLEVER$^{-EWC}$, showing that constructing pairs of pseudo queries and docids and supplementing them during continual indexing contributes to preventing forgetting for the retrieval ability. 
\item CLEVER$^{Random}$ shows the worst performance. Randomly selected old documents do not provide insights into the new documents and may introduce noise for continual indexing. 
\item CLEVER$^{-EWC}$ performs worse than CLEVER, showing the effectiveness of limiting the scope of model updates. 
\end{enumerate*}

Finally, CLEVER achieves the best performance.    
The results imply that applying an adaptive update strategy for PQ codes can assign effective docids to new documents without changing the old docids.
And rehearsing old similar documents and generating pseudo queries, can actively absorb knowledge from new documents while preserving previously learned retrieval ability. 

\begin{table}[t]
\small
    \caption{Model performance in non-incremental scenarios. We train on $\mathcal{D}_0$ and evaluate the performance of $\mathcal{Q}^\mathit{test}$. Note that DSI++ trained on $\mathcal{D}_0$ is DSI. $*$ indicates statistically significant improvements over all baselines (p-value $< 0.05$). }
    \label{tab:d0-res}
    \centering
    \setlength{\tabcolsep}{11pt}
    \renewcommand{\arraystretch}{0.8}
    \begin{tabular}{l@{ \ } ccc}
        \toprule
        \textbf{Model} & \textbf{CDI-MS (MRR@10)} & \textbf{CDI-NQ (Hits@10)} \\
        \midrule
        BM25 & 28.47 & 35.65 \\
        DPR & 38.51 & 40.37 \\
        DSI & 42.54 & 66.05 \\
        DSI-QG & 42.62 & 66.80 \\
        NCI & 43.14 & 67.49 \\
        Ultron & 42.78 & 67.13 \\
        CLEVER & \textbf{44.96}\rlap{$^*$} & \textbf{68.74}\rlap{$^*$} \\
        \bottomrule
    \end{tabular}
    \vspace*{-3mm}
\end{table}

\textbf{Incremental performance on a sequential query set}. The performance comparisons between CLEVER and the baselines on the \textit{sequential query set} are shown in Table~\ref{tab:seq-res} (recall that the metrics were introduced in Section~\ref{sec:formulation}). 
The relative order of different models under this setting in terms of \emph{VERT} is quite consistent with that on the previous setting of a single query set. 
For the evaluation metrics that measure the performance across different sessions in terms of AP, BWT and FWT, the full version of CLEVER achieves the best performance, again demonstrating the effectiveness of the proposed IPQ and learning mechanisms. 

\begin{figure*}[t]
 \centering
 \includegraphics[width=\textwidth]{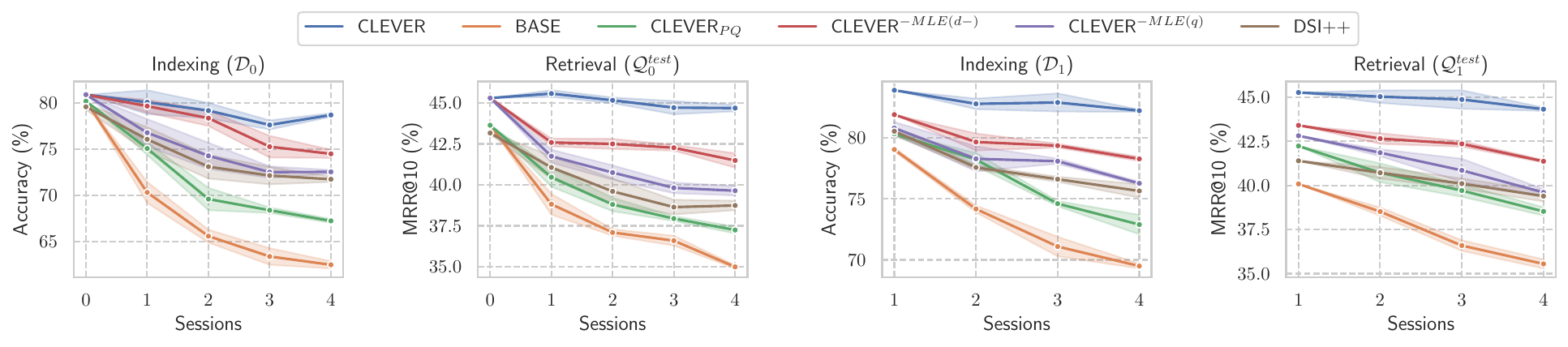}
 \caption{The catastrophic forgetting phenomenon of GR models. Based on the CDI-MS dataset, we illustrate the indexing accuracy of $\mathcal{D}_0$ and $\mathcal{D}_1$, and the retrieval MRR@10 of $\mathcal{Q}_0^\mathit{test}$ and $\mathcal{Q}_1^\mathit{test}$ under sequential query set setting.}
 \label{fig:forgetting}
 \vspace*{-4mm}
\end{figure*}

\textbf{Non-incremental performance}. To assess the performance of CLEVER before getting into the incremental aspect, we evaluate the performance of $\mathcal{Q}^\mathit{test}$ on $\mathcal{D}_0$ under the single query set setting.  
As shown in Table~\ref{tab:d0-res}, we see that: 
\begin{enumerate*}[label=(\roman*)]
	\item Compared with traditional IR models, CLEVER and existing generative retrieval methods achieve better performance, indicating the effectiveness of integrating different components into a single consolidated model.  
	\item CLEVER achieves better results than existing generative retrieval models, demonstrating that the two-step iterative process to learn discriminative PQ codes as docids  contributes to the retrieval effectiveness. 
\end{enumerate*}

\vspace*{-2mm}
\subsection{Assessing catastrophic forgetting}

To assess catastrophic forgetting of proposed methods, we show how the performance of the base session $\mathcal{D}_0$ and first incremental learning session $\mathcal{D}_1$ varies over the training process on the remaining sessions.   
For the indexing task, we evaluate the overall indexing accuracy of $\mathcal{D}_0$ and $\mathcal{D}_1$, i.e., we take a document as input of the GR model, if the generated sequence exactly matches with the correct docid, we treat the document as a positive sample. Otherwise, the document is a negative sample. 
For the retrieval task, we evaluate the performance of  $\mathcal{Q}_0^\mathit{test}$ and  $\mathcal{Q}_1^\mathit{test}$, i.e., MRR@10. 
See Figure~\ref{fig:forgetting}.

We observe that:  
\begin{enumerate*}[label=(\roman*)]  
\item CLEVER$_{PQ}$, CLEVER$^{-MLE(d-)}$ as well as CLEVER$^{-MLE(q)}$ suffer from catastrophic forgetting. 
Applying IPQ and the memory-augmented learning mechanism separately does not provide sufficient assurance for the model to perform well during continual document learning. 
\item DSI++ underperforms CLEVER by a large margin. A possible reason is that the atomic integers used in DSI++ as docids are difficult to quickly adapt to new documents and have a large impact on the docids of old documents, which may result in the loss of previously learned knowledge. 
\item Compared to CLEVER, the phenomenon of catastrophic forgetting is not as well mitigated in CLEVER$^{-MLE(d-)}$ and CLEVER$^{-MLE(q)}$, which underlines the importance of rehearsing old documents and the generated pseudo-queries. And
\item CLE\-VER almost avoids catastrophic forgetting on both indexing and retrieval tasks, showing its effectiveness in a dynamic setting.
\end{enumerate*} 

\begin{table}[t]
    \caption{Forward transfer analysis on CDI-MS under sequential query set. We evaluate the performance of $\mathcal{Q}^\mathit{test}_0$--$\mathcal{Q}^\mathit{test}_4$ in terms of VERT (\%), respectively. $*$ indicates statistically significant improvements over all baselines (p-value $< 0.05$).}
    \label{tab:forward}
    \centering
    \setlength{\tabcolsep}{9pt}
    \renewcommand{\arraystretch}{0.6}
    \begin{tabular}{l@{ \ } ccccc}
        \toprule
        \textbf{Model} & \multicolumn{1}{c}{\textbf{$\mathcal{D}_0$}} & \multicolumn{1}{c}{\textbf{$\mathcal{D}_1$}} & \multicolumn{1}{c}{\textbf{$\mathcal{D}_2$}} & \multicolumn{1}{c}{\textbf{$\mathcal{D}_3$}} & \multicolumn{1}{c}{\textbf{$\mathcal{D}_4$}} \\
        \midrule
      	INDIVIDUAL & 40.72 & 39.38 & 40.61 & 38.57 & 38.04 \\
      	CLEVER$_\mathit{init}$ & 45.30 & 40.63 & 41.15 & 40.94 & 40.22 \\
		\midrule
		CLEVER & \textbf{45.30} & \textbf{45.26}\rlap{$^*$} & \textbf{45.09}\rlap{$^*$} & \textbf{44.85}\rlap{$^*$} & \textbf{44.71}\rlap{$^*$}  \\ 
        \bottomrule
    \end{tabular}
\end{table}

\vspace*{-3mm}
\subsection{Assessing forward transfer}
Positive forward knowledge transfer is an essential ability during continual document learning including indexing and retrieval.  
Therefore, in this section, we explore the forward transfer ability of the CLEVER model, i.e.,  transferring knowledge from old documents to new documents.  
For CLEVER without initialization from the previous sessions we write CLEVER$_\mathit{init}$ and for individually fine-tuning the \ac{GR} model on each session we write INDIVIDUAL. 
Table \ref{tab:forward} displays the results. We see that:
\begin{enumerate*}[label=(\roman*)]
	\item CLEVER consistently and significantly outperforms INDIVIDUAL in the last four sessions. The underlying reason may be that CLEVER transfers old knowledge to new settings when continuously indexing new documents. INDIVIDUAL learns the indexing and retrieval tasks from each new session independently, which has a small size of data. 
	\item The performance improvements of CLEVER over CLEVER$_\mathit{init}$ further demonstrate the need for prior initialization in CLEVER, i.e., initializing new model parameters from the last parameters. 
\end{enumerate*}

\vspace*{-3mm}
\subsection{Effectiveness-efficiency trade-off} \label{sec:cost}
\begin{figure}[t]
 \centering
 \vspace*{-2mm}
 \includegraphics[width=0.48\textwidth]{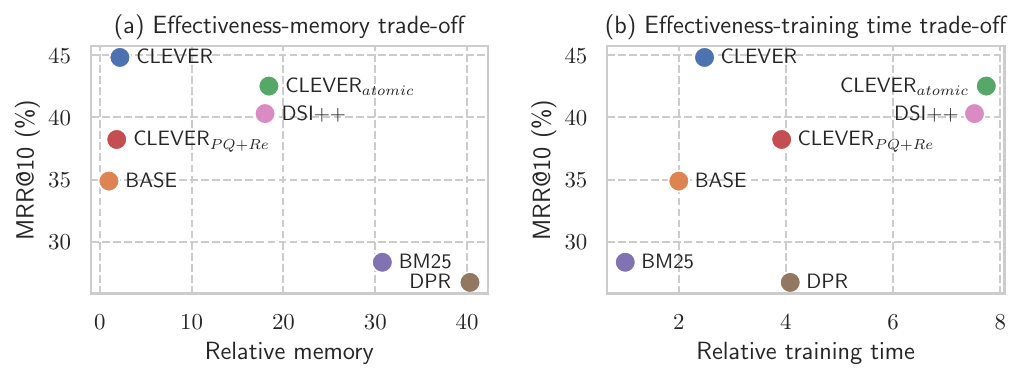}
 \caption{Comparison on (a) effectiveness-memory trade-off and (b) effectiveness-training time trade-off. Up and left is better. The search is performed on GPU.}
 \label{fig:cost}
\vspace*{-2mm}
\end{figure}

We evaluate the effectiveness and efficiency of different models on $\mathcal{Q}_{0}^\mathit{test}$ in terms of VERT(\%) after training all sessions. 
Regarding training time, we compare the overall training time by the end of the sequential training. 
For memory footprint, we compute the disk space occupied by the model at the training end. 
Figure~\ref{fig:cost} shows the relative memory and training time, i.e., the memory ratio and the time ratio of these methods with respect to BASE, respectively. 

Figure \ref{fig:cost} (a) shows the effectiveness-memory trade-off. We find that:
\begin{enumerate*}[label=(\roman*)]
    \item Traditional IR models (BM25 and DPR) consume much larger memory footprints than generative IR models, without discernible advantages in retrieval performance.  
   This result indicates the significant memory consumption of re-computing representations  and re-indexing for new documents.   
    \item Although CLEVER$_\mathit{atomic}$ and DSI++ are much more effective than the BASE model and CLEVER$_{PQ+Re}$, they suffer from severe memory inefficiency since they need the large softmax output space that comes with atomic docids and the embedding for each individual docid must be added to the model as new documents arrive. 
    \item CLEVER performs best in effectiveness and is almost as efficient as the BASE model. CLEVER only occupies a small amount of additional memory compared to BASE, which does not grow over sessions.  
\end{enumerate*}

Figure \ref{fig:cost} (b) shows the effectiveness-training time trade-off. We observe that:
\begin{enumerate*}[label=(\roman*)]
    \item BM25 exhibits a swift training process. However, its performance may be deemed suboptimal due to its vulnerability to the vocabulary mismatch issue, as well as its inability to adequately encapsulate semantic information. 
    \item The BASE model achieves training acceleration at the cost of compromised performance, which suggests that maintaining effectiveness is a non-trivial challenge for \ac{GR} in dynamic corpora setting. 
    \item CLEVER$_\mathit{atomic}$ and DSI++ sacrifice training time for effectiveness since the randomly initial embeddings require training from scratch.
    \item CLEVER$_{PQ+Re}$ re-trains all the centroids every session, leading to some computational overhead. However, the performance is still not improved too much.
    \item  CLEVER performs best in effectiveness and requires similar training times as the BASE model. 
\end{enumerate*}
These results demonstrate that CLEVER can be well deployed in practical environments due to its high effectiveness and efficiency.

\vspace*{-4mm}
\section{Related Work}

\textbf{Generative retrieval.} 
Recently, \acf{GR} has been proposed as an alternative paradigm \cite{metzler2021rethinking} for IR.  
Unlike the traditional ``index-then-rank'' paradigm \cite{bm25, zhan2021optimizing, dpr, mtdpr, kgi, guo2016deep, bert, prop, bprop}, a single seq2seq model is used to directly map a query to a relevant docid. 
In this way, the large external index is transferred into an internal index (i.e., model parameters), simplifying the data structure for supporting the retrieval process. 
Besides, it enables the end-to-end optimization towards the global objective for IR tasks. 

\ac{GR} is related to two key issues:
\begin{enumerate*}[label=(\roman*)]
\item the \textit{indexing task}: how to associate the content of each document with its docid, and 
\item the \textit{retrieval task}: how to map the queries to their relevant docids. 
\end{enumerate*} 
For the indexing task, previous efforts can boil down to two research lines. 
One is to generate docids for documents \cite{zhao2022dense} including atomic identifiers (e.g., a unique integer identifier \cite{dsi}), simple string identifiers (e.g., titles \cite{genre, gere, tang2023semantic}, n-grams \cite{seal,chen2023unified} or URLs \cite{zhou2022ultron}) and semantically structured identifiers (e.g., clustering-based prefix representation \cite{dsi} or PQ \cite{zhou2022ultron}). 
The other is to establish a semantic mapping from documents to docids. 
Various kinds of document content have been proposed to enhance the association \cite{dsi, zhou2022ultron, corpusbrain}, e.g., contexts at different semantic granularities  \cite{corpusbrain, zhou2022dynamicretriever} and hyperlink information  \cite{corpusbrain}. 
For the retrieval task, most approaches directly learn to map queries to relevant docids in an autoregressive way. 
Recently, some work has been adopted to generate pseudo-queries \cite{nci, zhou2022ultron} and designed pre-training tasks \cite{corpusbrain} to tackle the limited availability of labeled data. 
However, current \ac{GR} methods mainly focus on a stationary corpus scenario, i.e., with a fixed document collection. 

Very recently, \citet{mehta2022dsi++} have shown that continually memorizing new documents leads to considerable forgetting of old documents.
They directly assigned each new document an arbitrary unique integer identifier, and randomly  sampled some old documents using experience replay \cite{chaudhry2019tiny} for incremental updates. 
However, learning embeddings for each individual new docid from scratch incurs prohibitively high computational costs, while the relationships between new and old documents may not be easily obtained from randomly-selected exemplars. 
And they only considered the sequential query set setting for performance evaluation. 

In this work, the proposed IPQ technique is able to effectively represent new documents by updating a subset of centroids instead of all centroids, eliminating the need to update existing data indices. 
In IPQ, the partial codebook update strategy can be applied to other clustering-based docids, e.g., clustering-based prefix representation in DSI~\cite{dsi}, which we leave as future work.


\heading{Continual learning} \Acf{CL} has been a long-standing research topic to overcome the catastrophic forgetting problem of previously acquired knowledge, while continuously learning new knowledge from few labeled samples \cite{tan2022graph}. 
Recently, \ac{CL} has been considered in computer vision \cite{tao2020few,icarl} and natural language processing \cite{houlsby2019parameter,aljundi2017expert}, but few efforts have been devoted to IR so far.  
\ac{CL} scenarios \cite{mai2022online,ramesh2021model} can be divided into task increment, domain increment and class increment.  
In this work, we consider the practical setting of dynamic corpora with newly added  documents.  

Existing \ac{CL} approaches \cite{de2021continual} can be categorized into: 
\begin{enumerate*}[label=(\roman*)]
\item replay methods, maintaining a subset of previous samples and training models together with samples in the new session \cite{hinton2015distilling,zhao2020maintaining};
\item regularization-based methods, regularizing the model parameters to enable important parameters concerning the previous tasks to be protected when training on each new task \cite{ewc, zenke2017continual}; and 
\item parameter-isolation methods, dynamically allocating a set of parameters for each task \cite{aljundi2017expert}. 
\end{enumerate*}
Here, we take advantage of replay methods and regularization-based methods to memorize new documents. 

\vspace*{-1mm}
\section{Conclusion}

In this work, we have focused on a critical requirement for \acf{GR} models to be usable in practical scenarios, where new documents are continuously added to the corpus.   
In particular, we have presented a continual learning method to alleviate possible high computational costs for generating new docids, and leverage both past similar documents and pseudo-queries for consolidating knowledge. 
Extensive experiments have demonstrated the effectiveness and efficiency of our method. 

Despite the promising results that \ac{GR} has shown, its scalability remains a challenging issue, particularly concerning document addition, removal, and updates. These factors significantly impact the widespread adoption of \ac{GR} in various applications.
For the proposed CLEVER method, exploring the joint optimization of quantization methods in IPQ and \ac{GR} models using supervised labels, and devising advanced thresholds for adaptively updating PQ codes, hold great potential for enhancing retrieval effectiveness.


\vspace{-2mm}
\begin{acks}
    This work was funded by the National Natural Science Foundation of China (NSFC) under Grants No. 62006218 and 61902381, the China Scholarship Council under Grants No. 202104910234, the Youth Innovation Promotion Association CAS under Grants No. 2021100, the project under Grants No. JCKY2022130C039 and 2021QY1701,  the CAS Project for Young Scientists in Basic Research under Grant No. YSBR-034, the Innovation Project of ICT CAS under Grants No. E261090, and the Lenovo-CAS Joint Lab Youth Scientist Project.
    This work was also (partially) funded by the Hybrid Intelligence Center, a 10-year program funded by the Dutch Ministry of Education, Culture and Science through the Netherlands Organisation for Scientific Research, \url{https://hybrid-intelligence-centre.nl}, and project LESSEN with project number NWA.1389.20.183 of the research program NWA ORC 2020/21, which is (partly) financed by the Dutch Research Council (NWO).
    All content represents the opinion of the authors, which is not necessarily shared or endorsed by their respective employers and/or sponsors.
\end{acks}

\clearpage
\bibliographystyle{ACM-Reference-Format}
\bibliography{main}


\begin{thebibliography}{56}


\ifx \showCODEN    \undefined \def \showCODEN     #1{\unskip}     \fi
\ifx \showDOI      \undefined \def \showDOI       #1{#1}\fi
\ifx \showISBNx    \undefined \def \showISBNx     #1{\unskip}     \fi
\ifx \showISBNxiii \undefined \def \showISBNxiii  #1{\unskip}     \fi
\ifx \showISSN     \undefined \def \showISSN      #1{\unskip}     \fi
\ifx \showLCCN     \undefined \def \showLCCN      #1{\unskip}     \fi
\ifx \shownote     \undefined \def \shownote      #1{#1}          \fi
\ifx \showarticletitle \undefined \def \showarticletitle #1{#1}   \fi
\ifx \showURL      \undefined \def \showURL       {\relax}        \fi
\providecommand\bibfield[2]{#2}
\providecommand\bibinfo[2]{#2}
\providecommand\natexlab[1]{#1}
\providecommand\showeprint[2][]{arXiv:#2}

\bibitem[Aljundi et~al\mbox{.}(2017)]%
        {aljundi2017expert}
\bibfield{author}{\bibinfo{person}{Rahaf Aljundi}, \bibinfo{person}{Punarjay Chakravarty}, {and} \bibinfo{person}{Tinne Tuytelaars}.} \bibinfo{year}{2017}\natexlab{}.
\newblock \showarticletitle{Expert Gate: Lifelong Learning with a Network of Experts}. In \bibinfo{booktitle}{\emph{CVPR}}. \bibinfo{pages}{3366--3375}.
\newblock


\bibitem[Bevilacqua et~al\mbox{.}(2022)]%
        {seal}
\bibfield{author}{\bibinfo{person}{Michele Bevilacqua}, \bibinfo{person}{Giuseppe Ottaviano}, \bibinfo{person}{Patrick Lewis}, \bibinfo{person}{Wen tau Yih}, \bibinfo{person}{Sebastian Riedel}, {and} \bibinfo{person}{Fabio Petroni}.} \bibinfo{year}{2022}\natexlab{}.
\newblock \showarticletitle{Autoregressive Search Engines: Generating Substrings as Document Identifiers}. In \bibinfo{booktitle}{\emph{NeurIPS}}.
\newblock


\bibitem[Chaudhry et~al\mbox{.}(2019)]%
        {chaudhry2019tiny}
\bibfield{author}{\bibinfo{person}{Arslan Chaudhry}, \bibinfo{person}{Marcus Rohrbach}, \bibinfo{person}{Mohamed Elhoseiny}, \bibinfo{person}{Thalaiyasingam Ajanthan}, \bibinfo{person}{Puneet~K Dokania}, \bibinfo{person}{Philip~HS Torr}, {and} \bibinfo{person}{Marc'Aurelio Ranzato}.} \bibinfo{year}{2019}\natexlab{}.
\newblock \showarticletitle{On Tiny Episodic Memories in Continual Learning}.
\newblock \bibinfo{journal}{\emph{arXiv preprint arXiv:1902.10486}} (\bibinfo{year}{2019}).
\newblock


\bibitem[Chen et~al\mbox{.}(2023)]%
        {chen2023unified}
\bibfield{author}{\bibinfo{person}{Jiangui Chen}, \bibinfo{person}{Ruqing Zhang}, \bibinfo{person}{Jiafeng Guo}, \bibinfo{person}{Maarten de Rijke}, \bibinfo{person}{Yiqun Liu}, \bibinfo{person}{Yixing Fan}, {and} \bibinfo{person}{Xueqi Cheng}.} \bibinfo{year}{2023}\natexlab{}.
\newblock \showarticletitle{A Unified Generative Retriever for Knowledge-Intensive Language Tasks via Prompt Learning}. In \bibinfo{booktitle}{\emph{SIGIR}}.
\newblock


\bibitem[Chen et~al\mbox{.}(2022a)]%
        {gere}
\bibfield{author}{\bibinfo{person}{Jiangui Chen}, \bibinfo{person}{Ruqing Zhang}, \bibinfo{person}{Jiafeng Guo}, \bibinfo{person}{Yixing Fan}, {and} \bibinfo{person}{Xueqi Cheng}.} \bibinfo{year}{2022}\natexlab{a}.
\newblock \showarticletitle{GERE: Generative Evidence Retrieval for Fact Verification}. In \bibinfo{booktitle}{\emph{SIGIR}}. \bibinfo{pages}{2184--2189}.
\newblock


\bibitem[Chen et~al\mbox{.}(2022b)]%
        {corpusbrain}
\bibfield{author}{\bibinfo{person}{Jiangui Chen}, \bibinfo{person}{Ruqing Zhang}, \bibinfo{person}{Jiafeng Guo}, \bibinfo{person}{Yiqun Liu}, \bibinfo{person}{Yixing Fan}, {and} \bibinfo{person}{Xueqi Cheng}.} \bibinfo{year}{2022}\natexlab{b}.
\newblock \showarticletitle{CorpusBrain: Pre-train a Generative Retrieval Model for Knowledge-Intensive Language Tasks}. In \bibinfo{booktitle}{\emph{CIKM}}. \bibinfo{pages}{191--200}.
\newblock


\bibitem[Chen et~al\mbox{.}(2020)]%
        {Chen2020ASF}
\bibfield{author}{\bibinfo{person}{Ting Chen}, \bibinfo{person}{Simon Kornblith}, \bibinfo{person}{Mohammad Norouzi}, {and} \bibinfo{person}{Geoffrey~E. Hinton}.} \bibinfo{year}{2020}\natexlab{}.
\newblock \showarticletitle{A Simple Framework for Contrastive Learning of Visual Representations}. In \bibinfo{booktitle}{\emph{ICML}}, Vol.~\bibinfo{volume}{119}. \bibinfo{publisher}{{PMLR}}, \bibinfo{pages}{1597--1607}.
\newblock


\bibitem[Chen and He(2021)]%
        {Chen2021ExploringSS}
\bibfield{author}{\bibinfo{person}{Xinlei Chen} {and} \bibinfo{person}{Kaiming He}.} \bibinfo{year}{2021}\natexlab{}.
\newblock \showarticletitle{Exploring Simple Siamese Representation Learning}.
\newblock \bibinfo{journal}{\emph{CVPR}} (\bibinfo{year}{2021}), \bibinfo{pages}{15745--15753}.
\newblock


\bibitem[Clark et~al\mbox{.}(1977)]%
        {clark1977discourse}
\bibfield{author}{\bibinfo{person}{Herbert~H Clark}, \bibinfo{person}{S Haviland}, {and} \bibinfo{person}{Roy~O Freedle}.} \bibinfo{year}{1977}\natexlab{}.
\newblock \bibinfo{title}{Discourse Production and Comprehension}.
\newblock
\newblock


\bibitem[Clark and Haviland(1974)]%
        {clark1974psychological}
\bibfield{author}{\bibinfo{person}{Herbert~H Clark} {and} \bibinfo{person}{Susan~E Haviland}.} \bibinfo{year}{1974}\natexlab{}.
\newblock \showarticletitle{Psychological Processes as Linguistic Explanation}.
\newblock \bibinfo{journal}{\emph{Explaining linguistic phenomena}} (\bibinfo{year}{1974}), \bibinfo{pages}{91--124}.
\newblock


\bibitem[Danielsson(1980)]%
        {danielsson1980euclidean}
\bibfield{author}{\bibinfo{person}{Per-Erik Danielsson}.} \bibinfo{year}{1980}\natexlab{}.
\newblock \showarticletitle{Euclidean Distance Mapping}.
\newblock \bibinfo{journal}{\emph{CGIP}} (\bibinfo{year}{1980}).
\newblock


\bibitem[De~Cao et~al\mbox{.}(2020)]%
        {genre}
\bibfield{author}{\bibinfo{person}{Nicola De~Cao}, \bibinfo{person}{Gautier Izacard}, \bibinfo{person}{Sebastian Riedel}, {and} \bibinfo{person}{Fabio Petroni}.} \bibinfo{year}{2020}\natexlab{}.
\newblock \showarticletitle{Autoregressive Entity Retrieval}. In \bibinfo{booktitle}{\emph{ICLR}}.
\newblock


\bibitem[De~Lange et~al\mbox{.}(2021)]%
        {de2021continual}
\bibfield{author}{\bibinfo{person}{Matthias De~Lange}, \bibinfo{person}{Rahaf Aljundi}, \bibinfo{person}{Marc Masana}, \bibinfo{person}{Sarah Parisot}, \bibinfo{person}{Xu Jia}, \bibinfo{person}{Ale{\v{s}} Leonardis}, \bibinfo{person}{Gregory Slabaugh}, {and} \bibinfo{person}{Tinne Tuytelaars}.} \bibinfo{year}{2021}\natexlab{}.
\newblock \showarticletitle{A Continual Learning Survey: Defying Forgetting in Classification Tasks}.
\newblock \bibinfo{journal}{\emph{PAMI}} \bibinfo{volume}{44}, \bibinfo{number}{7} (\bibinfo{year}{2021}), \bibinfo{pages}{3366--3385}.
\newblock


\bibitem[Devlin et~al\mbox{.}(2019)]%
        {bert}
\bibfield{author}{\bibinfo{person}{Jacob Devlin}, \bibinfo{person}{Ming-Wei Chang}, \bibinfo{person}{Kenton Lee}, {and} \bibinfo{person}{Kristina Toutanova}.} \bibinfo{year}{2019}\natexlab{}.
\newblock \showarticletitle{BERT: Pre-training of Deep Bidirectional Transformers for Language Understanding}. In \bibinfo{booktitle}{\emph{NAACL-HLT}}. \bibinfo{pages}{4171--4186}.
\newblock


\bibitem[Glass et~al\mbox{.}(2021)]%
        {kgi}
\bibfield{author}{\bibinfo{person}{Michael Glass}, \bibinfo{person}{Gaetano Rossiello}, \bibinfo{person}{Md~Faisal~Mahbub Chowdhury}, {and} \bibinfo{person}{Alfio Gliozzo}.} \bibinfo{year}{2021}\natexlab{}.
\newblock \showarticletitle{Robust Retrieval Augmented Generation for Zero-shot Slot Filling}. In \bibinfo{booktitle}{\emph{EMNLP 2021}}. \bibinfo{pages}{1939--1949}.
\newblock


\bibitem[Guo et~al\mbox{.}(2016)]%
        {guo2016deep}
\bibfield{author}{\bibinfo{person}{Jiafeng Guo}, \bibinfo{person}{Yixing Fan}, \bibinfo{person}{Qingyao Ai}, {and} \bibinfo{person}{W~Bruce Croft}.} \bibinfo{year}{2016}\natexlab{}.
\newblock \showarticletitle{A Deep Relevance Matching Model for Ad-hoc Retrieval}. In \bibinfo{booktitle}{\emph{CIKM}}. \bibinfo{pages}{55--64}.
\newblock


\bibitem[Haviland and Clark(1974)]%
        {haviland1974s}
\bibfield{author}{\bibinfo{person}{Susan~E Haviland} {and} \bibinfo{person}{Herbert~H Clark}.} \bibinfo{year}{1974}\natexlab{}.
\newblock \showarticletitle{What's New? Acquiring New Information as a Process in Comprehension}.
\newblock \bibinfo{journal}{\emph{JVLBA}} \bibinfo{volume}{13}, \bibinfo{number}{5} (\bibinfo{year}{1974}), \bibinfo{pages}{512--521}.
\newblock


\bibitem[Hinton et~al\mbox{.}(2015)]%
        {hinton2015distilling}
\bibfield{author}{\bibinfo{person}{Geoffrey Hinton}, \bibinfo{person}{Oriol Vinyals}, \bibinfo{person}{Jeff Dean}, {et~al\mbox{.}}} \bibinfo{year}{2015}\natexlab{}.
\newblock \showarticletitle{Distilling the Knowledge in a Neural Network}.
\newblock \bibinfo{journal}{\emph{stat}}  \bibinfo{volume}{1050} (\bibinfo{year}{2015}), \bibinfo{pages}{9}.
\newblock


\bibitem[Houlsby et~al\mbox{.}(2019)]%
        {houlsby2019parameter}
\bibfield{author}{\bibinfo{person}{Neil Houlsby}, \bibinfo{person}{Andrei Giurgiu}, \bibinfo{person}{Stanislaw Jastrzebski}, \bibinfo{person}{Bruna Morrone}, \bibinfo{person}{Quentin De~Laroussilhe}, \bibinfo{person}{Andrea Gesmundo}, \bibinfo{person}{Mona Attariyan}, {and} \bibinfo{person}{Sylvain Gelly}.} \bibinfo{year}{2019}\natexlab{}.
\newblock \showarticletitle{Parameter-efficient Transfer Learning for NLP}. In \bibinfo{booktitle}{\emph{ICML}}. \bibinfo{pages}{2790--2799}.
\newblock


\bibitem[Jegou et~al\mbox{.}(2010)]%
        {jegou2010product}
\bibfield{author}{\bibinfo{person}{Herve Jegou}, \bibinfo{person}{Matthijs Douze}, {and} \bibinfo{person}{Cordelia Schmid}.} \bibinfo{year}{2010}\natexlab{}.
\newblock \showarticletitle{Product Quantization for Nearest Neighbor Search}.
\newblock \bibinfo{journal}{\emph{PAMI}} \bibinfo{volume}{33}, \bibinfo{number}{1} (\bibinfo{year}{2010}), \bibinfo{pages}{117--128}.
\newblock


\bibitem[Karpukhin et~al\mbox{.}(2020)]%
        {dpr}
\bibfield{author}{\bibinfo{person}{Vladimir Karpukhin}, \bibinfo{person}{Barlas Oguz}, \bibinfo{person}{Sewon Min}, \bibinfo{person}{Patrick Lewis}, \bibinfo{person}{Ledell Wu}, \bibinfo{person}{Sergey Edunov}, \bibinfo{person}{Danqi Chen}, {and} \bibinfo{person}{Wen-tau Yih}.} \bibinfo{year}{2020}\natexlab{}.
\newblock \showarticletitle{Dense Passage Retrieval for Open-Domain Question Answering}. In \bibinfo{booktitle}{\emph{EMNLP}}. \bibinfo{pages}{6769--6781}.
\newblock


\bibitem[Kingma and Ba(2015)]%
        {adam}
\bibfield{author}{\bibinfo{person}{Diederik~P. Kingma} {and} \bibinfo{person}{Jimmy Ba}.} \bibinfo{year}{2015}\natexlab{}.
\newblock \showarticletitle{Adam: {A} Method for Stochastic Optimization}. In \bibinfo{booktitle}{\emph{ICLR}}.
\newblock


\bibitem[Kirkpatrick et~al\mbox{.}(2017)]%
        {ewc}
\bibfield{author}{\bibinfo{person}{James Kirkpatrick}, \bibinfo{person}{Razvan Pascanu}, \bibinfo{person}{Neil Rabinowitz}, \bibinfo{person}{Joel Veness}, \bibinfo{person}{Guillaume Desjardins}, \bibinfo{person}{Andrei~A Rusu}, \bibinfo{person}{Kieran Milan}, \bibinfo{person}{John Quan}, \bibinfo{person}{Tiago Ramalho}, \bibinfo{person}{Agnieszka Grabska-Barwinska}, \bibinfo{person}{Demis Hassabis}, \bibinfo{person}{Claudia Clopath}, \bibinfo{person}{Dharshan Kumaran}, {and} \bibinfo{person}{Raia Hadsell}.} \bibinfo{year}{2017}\natexlab{}.
\newblock \showarticletitle{Overcoming Catastrophic Forgetting in Neural Networks}.
\newblock \bibinfo{journal}{\emph{Proceedings of the National Academy of Sciences}} \bibinfo{volume}{114}, \bibinfo{number}{13} (\bibinfo{year}{2017}).
\newblock


\bibitem[Kousha and Thelwall(2020)]%
        {kousha-2020-covid-19}
\bibfield{author}{\bibinfo{person}{Kayvan Kousha} {and} \bibinfo{person}{Mike Thelwall}.} \bibinfo{year}{2020}\natexlab{}.
\newblock \showarticletitle{{COVID-19} Publications: Database Coverage, Citations, Readers, Tweets, News, Facebook Walls, Reddit Posts}.
\newblock \bibinfo{journal}{\emph{Quantitative Science Studies}} \bibinfo{volume}{1}, \bibinfo{number}{3} (\bibinfo{year}{2020}), \bibinfo{pages}{1068--1091}.
\newblock


\bibitem[Kwiatkowski et~al\mbox{.}(2019)]%
        {nq}
\bibfield{author}{\bibinfo{person}{Tom Kwiatkowski}, \bibinfo{person}{Jennimaria Palomaki}, \bibinfo{person}{Olivia Redfield}, \bibinfo{person}{Michael Collins}, \bibinfo{person}{Ankur Parikh}, \bibinfo{person}{Chris Alberti}, \bibinfo{person}{Danielle Epstein}, \bibinfo{person}{Illia Polosukhin}, \bibinfo{person}{Jacob Devlin}, \bibinfo{person}{Kenton Lee}, {et~al\mbox{.}}} \bibinfo{year}{2019}\natexlab{}.
\newblock \showarticletitle{Natural Questions: a Benchmark for Question Answering Research}.
\newblock \bibinfo{journal}{\emph{Transactions of the Association for Computational Linguistics}}  \bibinfo{volume}{7} (\bibinfo{year}{2019}), \bibinfo{pages}{453--466}.
\newblock


\bibitem[Lopez-Paz and Ranzato(2017)]%
        {gem}
\bibfield{author}{\bibinfo{person}{David Lopez-Paz} {and} \bibinfo{person}{Marc'Aurelio Ranzato}.} \bibinfo{year}{2017}\natexlab{}.
\newblock \showarticletitle{Gradient Episodic Memory for Continual Learning}.
\newblock \bibinfo{journal}{\emph{NeurIPS}}  \bibinfo{volume}{30} (\bibinfo{year}{2017}).
\newblock


\bibitem[Ma et~al\mbox{.}(2022)]%
        {ma2022pre}
\bibfield{author}{\bibinfo{person}{Xinyu Ma}, \bibinfo{person}{Jiafeng Guo}, \bibinfo{person}{Ruqing Zhang}, \bibinfo{person}{Yixing Fan}, {and} \bibinfo{person}{Xueqi Cheng}.} \bibinfo{year}{2022}\natexlab{}.
\newblock \showarticletitle{Pre-train a Discriminative Text Encoder for Dense Retrieval via Contrastive Span Prediction}. In \bibinfo{booktitle}{\emph{SIGIR}}. \bibinfo{pages}{848–858}.
\newblock


\bibitem[Ma et~al\mbox{.}(2021a)]%
        {prop}
\bibfield{author}{\bibinfo{person}{Xinyu Ma}, \bibinfo{person}{Jiafeng Guo}, \bibinfo{person}{Ruqing Zhang}, \bibinfo{person}{Yixing Fan}, \bibinfo{person}{Xiang Ji}, {and} \bibinfo{person}{Xueqi Cheng}.} \bibinfo{year}{2021}\natexlab{a}.
\newblock \showarticletitle{PROP: Pre-training with Representative Words Prediction for Ad-hoc Retrieval}. In \bibinfo{booktitle}{\emph{WSDM}}. \bibinfo{pages}{283--291}.
\newblock


\bibitem[Ma et~al\mbox{.}(2021b)]%
        {bprop}
\bibfield{author}{\bibinfo{person}{Xinyu Ma}, \bibinfo{person}{Jiafeng Guo}, \bibinfo{person}{Ruqing Zhang}, \bibinfo{person}{Yixing Fan}, \bibinfo{person}{Yingyan Li}, {and} \bibinfo{person}{Xueqi Cheng}.} \bibinfo{year}{2021}\natexlab{b}.
\newblock \showarticletitle{B-PROP: Bootstrapped Pre-training with Representative Words Prediction for Ad-hoc Retrieval}. In \bibinfo{booktitle}{\emph{SIGIR}}. \bibinfo{pages}{1513--1522}.
\newblock


\bibitem[Mai et~al\mbox{.}(2022)]%
        {mai2022online}
\bibfield{author}{\bibinfo{person}{Zheda Mai}, \bibinfo{person}{Ruiwen Li}, \bibinfo{person}{Jihwan Jeong}, \bibinfo{person}{David Quispe}, \bibinfo{person}{Hyunwoo Kim}, {and} \bibinfo{person}{Scott Sanner}.} \bibinfo{year}{2022}\natexlab{}.
\newblock \showarticletitle{Online Continual Learning in Image Classification: An Empirical Survey}.
\newblock \bibinfo{journal}{\emph{Neurocomputing}}  \bibinfo{volume}{469} (\bibinfo{year}{2022}), \bibinfo{pages}{28--51}.
\newblock


\bibitem[Maillard et~al\mbox{.}(2021)]%
        {mtdpr}
\bibfield{author}{\bibinfo{person}{Jean Maillard}, \bibinfo{person}{Vladimir Karpukhin}, \bibinfo{person}{Fabio Petroni}, \bibinfo{person}{Wen-tau Yih}, \bibinfo{person}{Barlas Oguz}, \bibinfo{person}{Veselin Stoyanov}, {and} \bibinfo{person}{Gargi Ghosh}.} \bibinfo{year}{2021}\natexlab{}.
\newblock \showarticletitle{Multi-Task Retrieval for Knowledge-Intensive Tasks}. In \bibinfo{booktitle}{\emph{ACL}}. \bibinfo{pages}{1098--1111}.
\newblock


\bibitem[Manning et~al\mbox{.}(2008)]%
        {manning2008introduction}
\bibfield{author}{\bibinfo{person}{Christopher~D. Manning}, \bibinfo{person}{Prabhakar Raghavan}, {and} \bibinfo{person}{Hinrich Schütze}.} \bibinfo{year}{2008}\natexlab{}.
\newblock \bibinfo{booktitle}{\emph{Introduction to Information Retrieval}}.
\newblock \bibinfo{publisher}{Cambridge University Press}.
\newblock


\bibitem[Mehta et~al\mbox{.}(2022)]%
        {mehta2022dsi++}
\bibfield{author}{\bibinfo{person}{Sanket~Vaibhav Mehta}, \bibinfo{person}{Jai Gupta}, \bibinfo{person}{Yi Tay}, \bibinfo{person}{Mostafa Dehghani}, \bibinfo{person}{Vinh~Q Tran}, \bibinfo{person}{Jinfeng Rao}, \bibinfo{person}{Marc Najork}, \bibinfo{person}{Emma Strubell}, {and} \bibinfo{person}{Donald Metzler}.} \bibinfo{year}{2022}\natexlab{}.
\newblock \showarticletitle{DSI++: Updating Transformer Memory with New Documents}.
\newblock \bibinfo{journal}{\emph{arXiv preprint arXiv:2212.09744}} (\bibinfo{year}{2022}).
\newblock


\bibitem[Metzler et~al\mbox{.}(2021)]%
        {metzler2021rethinking}
\bibfield{author}{\bibinfo{person}{Donald Metzler}, \bibinfo{person}{Yi Tay}, \bibinfo{person}{Dara Bahri}, {and} \bibinfo{person}{Marc Najork}.} \bibinfo{year}{2021}\natexlab{}.
\newblock \showarticletitle{Rethinking Search: Making Domain Experts Out of Dilettantes}.
\newblock \bibinfo{journal}{\emph{ACM SIGIR Forum}} \bibinfo{volume}{55}, \bibinfo{number}{1} (\bibinfo{year}{2021}), \bibinfo{pages}{1--27}.
\newblock


\bibitem[Myung(2003)]%
        {mle}
\bibfield{author}{\bibinfo{person}{In~Jae Myung}.} \bibinfo{year}{2003}\natexlab{}.
\newblock \showarticletitle{Tutorial on Maximum Likelihood Estimation}.
\newblock \bibinfo{journal}{\emph{Journal of Mathematical Psychology}} \bibinfo{volume}{47}, \bibinfo{number}{1} (\bibinfo{year}{2003}), \bibinfo{pages}{90--100}.
\newblock


\bibitem[Nguyen et~al\mbox{.}(2016)]%
        {msmarco}
\bibfield{author}{\bibinfo{person}{Tri Nguyen}, \bibinfo{person}{Mir Rosenberg}, \bibinfo{person}{Xia Song}, \bibinfo{person}{Jianfeng Gao}, \bibinfo{person}{Saurabh Tiwary}, \bibinfo{person}{Rangan Majumder}, {and} \bibinfo{person}{Li Deng}.} \bibinfo{year}{2016}\natexlab{}.
\newblock \showarticletitle{MS MARCO: A Human Generated Machine Reading Comprehension Dataset}. In \bibinfo{booktitle}{\emph{CoCo@NIPs}}.
\newblock


\bibitem[Nogueira et~al\mbox{.}(2019)]%
        {doc2query}
\bibfield{author}{\bibinfo{person}{Rodrigo Nogueira}, \bibinfo{person}{Wei Yang}, \bibinfo{person}{Jimmy Lin}, {and} \bibinfo{person}{Kyunghyun Cho}.} \bibinfo{year}{2019}\natexlab{}.
\newblock \showarticletitle{Document Expansion by Query Prediction}.
\newblock \bibinfo{journal}{\emph{arXiv preprint arXiv:1904.08375}} (\bibinfo{year}{2019}).
\newblock


\bibitem[Raffel et~al\mbox{.}(2020)]%
        {t5}
\bibfield{author}{\bibinfo{person}{Colin Raffel}, \bibinfo{person}{Noam Shazeer}, \bibinfo{person}{Adam Roberts}, \bibinfo{person}{Katherine Lee}, \bibinfo{person}{Sharan Narang}, \bibinfo{person}{Michael Matena}, \bibinfo{person}{Yanqi Zhou}, \bibinfo{person}{Wei Li}, {and} \bibinfo{person}{Peter~J Liu}.} \bibinfo{year}{2020}\natexlab{}.
\newblock \showarticletitle{Exploring the Limits of Transfer Learning with a Unified Text-to-Text Transformer}.
\newblock \bibinfo{journal}{\emph{Journal of Machine Learning Research}}  \bibinfo{volume}{21} (\bibinfo{year}{2020}), \bibinfo{pages}{1--67}.
\newblock


\bibitem[Ramesh and Chaudhari(2021)]%
        {ramesh2021model}
\bibfield{author}{\bibinfo{person}{Rahul Ramesh} {and} \bibinfo{person}{Pratik Chaudhari}.} \bibinfo{year}{2021}\natexlab{}.
\newblock \showarticletitle{Model Zoo: A Growing Brain That Learns Continually}. In \bibinfo{booktitle}{\emph{ICLR}}.
\newblock


\bibitem[Rebuffi et~al\mbox{.}(2017)]%
        {icarl}
\bibfield{author}{\bibinfo{person}{Sylvestre-Alvise Rebuffi}, \bibinfo{person}{Alexander Kolesnikov}, \bibinfo{person}{Georg Sperl}, {and} \bibinfo{person}{Christoph~H Lampert}.} \bibinfo{year}{2017}\natexlab{}.
\newblock \showarticletitle{iCaRL: Incremental Classifier and Representation Learning}. In \bibinfo{booktitle}{\emph{CVPR}}. \bibinfo{pages}{2001--2010}.
\newblock


\bibitem[Robertson and Zaragoza(2009)]%
        {bm25}
\bibfield{author}{\bibinfo{person}{Stephen Robertson} {and} \bibinfo{person}{Hugo Zaragoza}.} \bibinfo{year}{2009}\natexlab{}.
\newblock \bibinfo{booktitle}{\emph{The Probabilistic Relevance Framework: BM25 and Beyond}}.
\newblock \bibinfo{publisher}{Now Publishers Inc}.
\newblock


\bibitem[Sutskever et~al\mbox{.}(2014)]%
        {sutskever2014sequence}
\bibfield{author}{\bibinfo{person}{Ilya Sutskever}, \bibinfo{person}{Oriol Vinyals}, {and} \bibinfo{person}{Quoc~V Le}.} \bibinfo{year}{2014}\natexlab{}.
\newblock \showarticletitle{Sequence to Sequence Learning with Neural Networks}.
\newblock \bibinfo{journal}{\emph{NeurIPS}} (\bibinfo{year}{2014}).
\newblock


\bibitem[Tan et~al\mbox{.}(2022)]%
        {tan2022graph}
\bibfield{author}{\bibinfo{person}{Zhen Tan}, \bibinfo{person}{Kaize Ding}, \bibinfo{person}{Ruocheng Guo}, {and} \bibinfo{person}{Huan Liu}.} \bibinfo{year}{2022}\natexlab{}.
\newblock \showarticletitle{Graph Few-shot Class-incremental Learning}. In \bibinfo{booktitle}{\emph{WSDM}}. \bibinfo{pages}{987--996}.
\newblock


\bibitem[Tang et~al\mbox{.}(2023)]%
        {tang2023semantic}
\bibfield{author}{\bibinfo{person}{Yubao Tang}, \bibinfo{person}{Ruqing Zhang}, \bibinfo{person}{Jiafeng Guo}, \bibinfo{person}{Jiangui Chen}, \bibinfo{person}{Zuowei Zhu}, \bibinfo{person}{Shuaiqiang Wang}, \bibinfo{person}{Dawei Yin}, {and} \bibinfo{person}{Xueqi Cheng}.} \bibinfo{year}{2023}\natexlab{}.
\newblock \showarticletitle{Semantic-Enhanced Differentiable Search Index Inspired by Learning Strategies}.
\newblock \bibinfo{journal}{\emph{arXiv preprint arXiv:2305.15115}} (\bibinfo{year}{2023}).
\newblock


\bibitem[Tao et~al\mbox{.}(2020)]%
        {tao2020few}
\bibfield{author}{\bibinfo{person}{Xiaoyu Tao}, \bibinfo{person}{Xiaopeng Hong}, \bibinfo{person}{Xinyuan Chang}, \bibinfo{person}{Songlin Dong}, \bibinfo{person}{Xing Wei}, {and} \bibinfo{person}{Yihong Gong}.} \bibinfo{year}{2020}\natexlab{}.
\newblock \showarticletitle{Few-shot Class-incremental Learning}. In \bibinfo{booktitle}{\emph{CVPR}}.
\newblock


\bibitem[Tay et~al\mbox{.}(2022)]%
        {dsi}
\bibfield{author}{\bibinfo{person}{Yi Tay}, \bibinfo{person}{Vinh~Q Tran}, \bibinfo{person}{Mostafa Dehghani}, \bibinfo{person}{Jianmo Ni}, \bibinfo{person}{Dara Bahri}, \bibinfo{person}{Harsh Mehta}, \bibinfo{person}{Zhen Qin}, \bibinfo{person}{Kai Hui}, \bibinfo{person}{Zhe Zhao}, \bibinfo{person}{Jai Gupta}, {et~al\mbox{.}}} \bibinfo{year}{2022}\natexlab{}.
\newblock \showarticletitle{Transformer Memory as a Differentiable Search Index}. In \bibinfo{booktitle}{\emph{NeurIPS}}.
\newblock


\bibitem[Wang et~al\mbox{.}(2022)]%
        {nci}
\bibfield{author}{\bibinfo{person}{Yujing Wang}, \bibinfo{person}{Yingyan Hou}, \bibinfo{person}{Haonan Wang}, \bibinfo{person}{Ziming Miao}, \bibinfo{person}{Shibin Wu}, \bibinfo{person}{Qi Chen}, \bibinfo{person}{Yuqing Xia}, \bibinfo{person}{Chengmin Chi}, \bibinfo{person}{Guoshuai Zhao}, \bibinfo{person}{Zheng Liu}, {et~al\mbox{.}}} \bibinfo{year}{2022}\natexlab{}.
\newblock \showarticletitle{A Neural Corpus Indexer for Document Retrieval}.
\newblock \bibinfo{journal}{\emph{Advances in Neural Information Processing Systems}}  \bibinfo{volume}{35} (\bibinfo{year}{2022}), \bibinfo{pages}{25600--25614}.
\newblock


\bibitem[Witten et~al\mbox{.}(2010)]%
        {witten-2010-how}
\bibfield{author}{\bibinfo{person}{Ian~H. Witten}, \bibinfo{person}{David Bainbridge}, {and} \bibinfo{person}{David~M. Nichols}.} \bibinfo{year}{2010}\natexlab{}.
\newblock \bibinfo{booktitle}{\emph{How to Build a Digital Library}}.
\newblock \bibinfo{publisher}{Morgan Kaufmann}.
\newblock


\bibitem[Xu et~al\mbox{.}(2022)]%
        {xu2022improving}
\bibfield{author}{\bibinfo{person}{Shicheng Xu}, \bibinfo{person}{Liang Pang}, \bibinfo{person}{Huawei Shen}, {and} \bibinfo{person}{Xueqi Cheng}.} \bibinfo{year}{2022}\natexlab{}.
\newblock \showarticletitle{Improving Multi-task Generalization Ability for Neural Text Matching via Prompt Learning}.
\newblock \bibinfo{journal}{\emph{arXiv preprint arXiv:2204.02725}} (\bibinfo{year}{2022}).
\newblock


\bibitem[Zenke et~al\mbox{.}(2017)]%
        {zenke2017continual}
\bibfield{author}{\bibinfo{person}{Friedemann Zenke}, \bibinfo{person}{Ben Poole}, {and} \bibinfo{person}{Surya Ganguli}.} \bibinfo{year}{2017}\natexlab{}.
\newblock \showarticletitle{Continual Learning through Synaptic Intelligence}. In \bibinfo{booktitle}{\emph{ICML}}. PMLR, \bibinfo{pages}{3987--3995}.
\newblock


\bibitem[Zhan et~al\mbox{.}(2021)]%
        {zhan2021optimizing}
\bibfield{author}{\bibinfo{person}{Jingtao Zhan}, \bibinfo{person}{Jiaxin Mao}, \bibinfo{person}{Yiqun Liu}, \bibinfo{person}{Jiafeng Guo}, \bibinfo{person}{Min Zhang}, {and} \bibinfo{person}{Shaoping Ma}.} \bibinfo{year}{2021}\natexlab{}.
\newblock \showarticletitle{Optimizing Dense Retrieval Model Training with Hard Negatives}. In \bibinfo{booktitle}{\emph{Proceedings of the 44th International ACM SIGIR Conference on Research and Development in Information Retrieval}}. \bibinfo{pages}{1503--1512}.
\newblock


\bibitem[Zhao et~al\mbox{.}(2020)]%
        {zhao2020maintaining}
\bibfield{author}{\bibinfo{person}{Bowen Zhao}, \bibinfo{person}{Xi Xiao}, \bibinfo{person}{Guojun Gan}, \bibinfo{person}{Bin Zhang}, {and} \bibinfo{person}{Shu-Tao Xia}.} \bibinfo{year}{2020}\natexlab{}.
\newblock \showarticletitle{Maintaining Discrimination and Fairness in Class Incremental Learning}. In \bibinfo{booktitle}{\emph{CVPR}}.
\newblock


\bibitem[Zhao et~al\mbox{.}(2022)]%
        {zhao2022dense}
\bibfield{author}{\bibinfo{person}{Wayne~Xin Zhao}, \bibinfo{person}{Jing Liu}, \bibinfo{person}{Ruiyang Ren}, {and} \bibinfo{person}{Ji-Rong Wen}.} \bibinfo{year}{2022}\natexlab{}.
\newblock \showarticletitle{Dense Text Retrieval Based on Pretrained Language Models: A Survey}.
\newblock \bibinfo{journal}{\emph{arXiv preprint arXiv:2211.14876}} (\bibinfo{year}{2022}).
\newblock


\bibitem[Zhou et~al\mbox{.}(2022)]%
        {zhou2022ultron}
\bibfield{author}{\bibinfo{person}{Yujia Zhou}, \bibinfo{person}{Jing Yao}, \bibinfo{person}{Zhicheng Dou}, \bibinfo{person}{Ledell Wu}, \bibinfo{person}{Peitian Zhang}, {and} \bibinfo{person}{Ji-Rong Wen}.} \bibinfo{year}{2022}\natexlab{}.
\newblock \showarticletitle{Ultron: An Ultimate Retriever on Corpus with a Model-based Indexer}.
\newblock \bibinfo{journal}{\emph{arXiv preprint arXiv:2208.09257}} (\bibinfo{year}{2022}).
\newblock


\bibitem[Zhou et~al\mbox{.}(2023)]%
        {zhou2022dynamicretriever}
\bibfield{author}{\bibinfo{person}{Yu-Jia Zhou}, \bibinfo{person}{Jing Yao}, \bibinfo{person}{Zhi-Cheng Dou}, \bibinfo{person}{Ledell Wu}, {and} \bibinfo{person}{Ji-Rong Wen}.} \bibinfo{year}{2023}\natexlab{}.
\newblock \showarticletitle{{DynamicRetriever: A Pre-trained Model-based IR System Without an Explicit Index}}.
\newblock \bibinfo{journal}{\emph{Machine Intelligence Research}} \bibinfo{volume}{20}, \bibinfo{number}{2} (\bibinfo{year}{2023}), \bibinfo{pages}{276--288}.
\newblock


\bibitem[Zhuang et~al\mbox{.}(2022)]%
        {zhuang2022bridging}
\bibfield{author}{\bibinfo{person}{Shengyao Zhuang}, \bibinfo{person}{Houxing Ren}, \bibinfo{person}{Linjun Shou}, \bibinfo{person}{Jian Pei}, \bibinfo{person}{Ming Gong}, \bibinfo{person}{Guido Zuccon}, {and} \bibinfo{person}{Daxin Jiang}.} \bibinfo{year}{2022}\natexlab{}.
\newblock \showarticletitle{Bridging the Gap between Indexing and Retrieval for Differentiable Search Index with Query Generation}.
\newblock \bibinfo{journal}{\emph{arXiv preprint arXiv:2206.10128}} (\bibinfo{year}{2022}).
\newblock


\end{thebibliography}


\end{document}